\documentclass[12pt, singlecolumn]{article}

\usepackage{arxiv}
\usepackage[utf8]{inputenc} 
\usepackage[T1]{fontenc}    
\usepackage{hyperref}       
\usepackage{url}            
\usepackage{amsmath}
\usepackage{amssymb}
\usepackage{booktabs}       
\usepackage{amsfonts}       
\usepackage{microtype}      
\usepackage{lipsum}		
\usepackage{graphicx}
\usepackage{natbib}
\usepackage{doi}

\title{Measure synchronization in interacting Hamiltonian systems: A brief review}

\date{} 					

\author{{Anupam Ghosh}\\
	Department of Complex Systems, Institute of Computer Science, Czech Academy of Sciences, \\Prague 18207, Czech Republic\\
	\texttt{anupamghosh0019@gmail.com} \\
}

\renewcommand{\headeright}{A. Ghosh}
\renewcommand{\undertitle}{}
\renewcommand{\shorttitle}{Measure synchronization in interacting Hamiltonian systems}

\begin{document}
\maketitle

\begin{abstract}
This paper aims to review the measure synchronization, a weak form of synchronization observed in coupled Hamiltonian systems, briefly. This synchronization is characterized by a Hamiltonian system that displays either quasiperiodic or chaotic dynamics. Each system, in the presence of either linear or nonlinear coupling, shares a phase space domain with an identical invariant measure in the measure synchronized state. It is important to note that while the trajectories are identical in measure, they do not necessarily exhibit complete temporal synchrony. This synchronization has been observed in various physical systems, such as coupled pendulums, Josephson junctions, and lasers.
\end{abstract}


\section{Introduction}
\label{sec:intro}
The word `synchronization' combines two Greek words: \emph{Syn} and \emph{Chronos}. These words, syn and chronos, imply common (or identical) and time, respectively. Synchronization in dynamical systems refers to a phenomenon where two or more systems evolve in a coordinated manner due to the interaction or coupling between them~\citep{pikovsky01,arenas08, balanov08, ghosh20_2}. This phenomenon is first observed in the late seventeenth century by C. Huygens~\citep{huygens73}. Huygens discovers that when two pendulum clocks are placed on the same wooden beam, they would eventually synchronize their oscillations, even if they have started with different initial conditions. Huygens explains this phenomenon through his theory of coupled oscillators, where the coupling force between the oscillators is transmitted through the wooden beam. Subsequent studies~\citep{pikovsky01, balanov08} report that synchronization can occur in various forms, such as phase synchronization, frequency synchronization, and complete synchronization, depending on the nature of the dynamical systems and the coupling between them. The study of synchronization in dynamical systems is of great interest in many fields, including physics, biology, neuroscience, and engineering, and has important applications in information processing, communication, and control~\citep{pecora97,arenas08, balanov08}.

Synchronization is observed in various dynamical systems, including coupled dissipative and Hamiltonian systems. A Hamiltonian system~\citep{lichtenberg92} is a conservative system that conserves the phase space volume with time following Liouville's theorem. The motion of the system is described by a set of differential equations, known as Hamilton's equations, that describe how the generalized coordinates and corresponding generalized momentum variables of the system evolve over time. The synchronization in Hamiltonian systems results from the exchange and conservation of energy between the interconnected components. This can be observed in systems like coupled pendulums or celestial bodies, where the shared energy leads to synchronized motion without any net loss. A dissipative system~\citep{strogatz07}, on the other hand, is a system that loses energy over time due to internal friction, viscosity, or other forms of energy dissipation. A set of differential equations also describes the motion of any dissipative system, but the equations include terms that represent the dissipation of energy. As a result, the phase space volume collapses with evolution time, and the system eventually reaches an attractor or a steady state. Overall, the dynamics of coupled dissipative and coupled Hamiltonian systems differ, and this dissimilarity leads to different types of synchronization in both categories. However, we restrict ourselves to study measure synchronization~\citep{hampton99}, observed in coupled Hamiltonian systems, in this review. 

The synchronization phenomenon in Hamiltonian systems has been observed in various physical systems, including coupled pendulums, Josephson junctions, and lasers~\citep{tian13,qiu14,qiu15,bemani17}. The understanding of synchronization in such systems has led to the development of several analytical and numerical techniques and has found applications in various fields. Therefore, studying synchronization in coupled Hamiltonian systems is a complex and fascinating topic that has attracted significant research interest. Measure synchronization (MS)~\citep{hampton99} is a phenomenon observed in coupled Hamiltonian systems where the distributions of the measures of two or more systems become synchronized, while the individual trajectories of each system may still vary widely. Here, the word `measure' refers to the Lebesgue measure~\citep{legesgue02}, a mathematical concept used to measure the size or volume of subsets of Euclidean space. The onset of MS implies that the statistical properties of the states of the systems are highly similar, but the specific values of their coordinates may differ significantly.

MS is observed in a coupled-pendulum system with a common beam serving as the coupling medium, resulting in indirect coupling between the pendulums~\citep{tian19}. This study~\citep{tian19} focuses on the role of the common beam in the emergence of MS and aims to contribute to the conventional understanding of synchronization (proposed by C. Huygens~\citep{huygens73}) by emphasizing the crucial role played by the common beam. The coupled-pendulum system suspended from the common beam can reach MS by adjusting the mass ratio of the common beam to that of each pendulum. Classical manifestations of MS have also been observed across various model systems, including the $\phi^4$ model~\citep{wang03}, Duffing model~\citep{vincent05}, and coupled bosonic Josephson junction~\citep{tian13}. The classical concept of MS has further been extended into quantum mechanics, and it has been demonstrated that quantum MS emerges in coupled quantum many-body systems~\citep{qiu14}. In the quantum domain, MS is characterized by purely quantum mechanical phenomena; specifically, collapses and revivals dynamics become synchronized once the system reaches a state of quantum MS. Thus, MS is observed in a wide range of physical systems, such as coupled oscillators, coupled maps, and coupled particles. To this end, the partial MS~\citep{wang02} manifests in systems comprising more than two suitably interconnected subsystems. This phenomenon is notably evident in systems involving three subsystems, wherein a distinct pattern emerges: the first subsystem may exhibit synchronization with the third subsystem while remaining uncoordinated with the second subsystem. This intriguing phenomenon is termed partial MS~\citep{vincent05,de18}.

The structure of this paper is organized as follows: firstly, we explore an analytical understanding of the fundamental mechanism underlying MS in Sec.~\ref{sec:ms}. Subsequently, we inspect two broad categories of MS transitions in Sec.~\ref{sec:sec_tran}. In this section, we also discuss various indices found in the existing literature, which facilitate the quantitative detection of MS within dynamical systems. Furthermore, we emphasize the significance of the initial condition selection in this context. Sec.~\ref{sec:pms} is dedicated to discussions surrounding partial MS. Sec.~\ref{sec:ocs} delves into the framework of occasional coupling as a tool for studying MS. Sec.~\ref{sec:quantum} investigates the exploration of MS within quantum Hamiltonian systems. Finally, in Sec.~\ref{sec:con}, we summarize MS and discuss different possible future aspects of MS in interacting dynamical systems.

\section{Mechanism of MS}
\label{sec:ms}
In this section, we analytically understand the underlying mechanism of MS, as reported by Hampton and Zanette~\citep{hampton99}. In this regard, we adopt the following example of coupled Hamiltonian systems~\citep{hampton99}:
\begin{equation}
	H =\frac{\omega_1^2}{2}+\frac{\omega_2^2}{2}- \frac{K}{2} \cos(\theta_2-\theta_1).\label{eq:sho}
\end{equation}
Here, two dynamical systems with Hamiltonian $H_i = \frac{\omega_i^2}{2}$, $i = 1, 2$, are interacting with each other through the term $\frac{K}{2} \cos(\theta_2-\theta_1)$. The strength of this interaction is measured by the parameter $K$. The equations of motion in canonical form can be expressed as:
\begin{subequations}
	\begin{eqnarray}
		\dot{\theta}_1&=&\omega_1, \\
		\dot{\theta}_2&=&\omega_2, \\
		\dot{\omega}_1&=&\frac{K}{2} \sin(\theta_2-\theta_1), \\
		\dot{\omega}_2&=&\frac{K}{2} \sin(\theta_1-\theta_2).
	\end{eqnarray}\label{eq:EOM_sho}
\end{subequations}
We begin with $K = 0$, i.e., when no interaction is activated between the Hamiltonian systems. In this case, the angle variable $\theta_i$ changes with a constant rate $\omega_i$, which further help us to write: 
\begin{equation}
	\omega_1(t) + \omega_2(t) = \omega_1(0) + \omega_2(0) = \Omega, 
\end{equation}
where $\Omega$ is a constant. The angle variables can also be written as:
\begin{equation}
	\theta_1(t) + \theta_2(t) = \theta_1(0) + \theta_2(0) + \Omega t.
\end{equation}
Now, introduce two new variables $\xi(t)$ and $\nu(t)$ as follows:
\begin{subequations}
	\begin{eqnarray}
		\xi(t) &=& \theta_1(t) - \theta_2(t), \\\omega_{1,2}(t) &=& \omega_{1,2}(0) \pm \nu(t).\label{eq:new_para}
	\end{eqnarray}
\end{subequations}
In terms of the newly defined variables, we can write:
\begin{subequations}
	\begin{eqnarray}
		\dot{\xi}(t) &=& \omega_0 + 2 \nu, \label{eq:6a}\\\dot{\nu}(t) &=& -\frac{K}{2} \sin\xi,\label{eq:6b}
	\end{eqnarray}
\end{subequations}
where $\omega_0 = \omega_1 (0) - \omega_2 (0)$. Differentiating Eq.~\ref{eq:6a} and with the help of Eq.~\ref{eq:6b}, we can write:
\begin{equation}
	\ddot{\xi}(t) + K \sin\xi = 0.
	\label{eq:sho}
\end{equation}
Thus, Eq.~\ref{eq:sho} represents a pendulum with total energy:  
\begin{subequations}
	\begin{eqnarray}
		E &=& \frac{\dot{\xi}^2(t)}{2} - K \cos\xi(t),\\ & =& \frac{\dot{\xi}^2(0)}{2} - K \cos\xi(0).
		\label{eq:energy_sho}
	\end{eqnarray}
\end{subequations}
The total energy $E$ remains invariant with the evolution time $t$. When $E > K$, $\xi(t)$ changes monotonically with time $t$, and $|\dot{\xi}|$ oscillates between $\sqrt{2(E-K)}$ and $\sqrt{2(E+K)}$. This kind of dynamics is termed `rotation'. In contrast, an oscillation of the pendulum is detected around the fixed point $\xi = 0$ for $E < K$. The variables $\xi$ and $\dot{\xi}$ evolve symmetrically around $\xi = 0$. These two different types of dynamics, rotations and oscillations, are separated by the separatrix $E=K$. The threshold value of the coupling strength $K^{c}$ is given by:
\begin{equation}
	K^c = \frac{\dot{\xi}^2(0)}{2\left(1 + \cos \xi(0)\right)},
	\label{eq:kth}
\end{equation}
with the total energy $E = K^{c} + (K^{c} - K) \cos \xi(0)$.
The original frequency variables become $\omega_{1,2} = (\Omega \pm \dot{\xi})/2$. For a significantly smaller value of $K$, $\dot{\xi}$ exhibits small oscillation around $\sqrt{E} \simeq \sqrt{2K^{c}}$, and frequencies $\omega_1$ and $\omega_2$ oscillate around two separated values $\Omega/2 + \sqrt{K^{c}/2}$ and $\Omega/2 - \sqrt{K^c/2}$, respectively. A further increase in $K$, as long as $K < K^c$, the frequencies evolve within two non-overlapping intervals $\left[\Omega/2 + \sqrt{(E-K)/2}, \Omega/2 + \sqrt{(E+K)/2}\right]$ and $\left[\Omega/2 - \sqrt{(E-K)/2}, \Omega/2 - \sqrt{(E+K)/2}\right]$. It is noteworthy that as the coupling strength $K$ approaches its threshold value, both the lower boundary of the upper interval and the upper boundary of the lower interval converge toward the value of $\Omega/2$. Finally, as soon as $K$ crosses $K^c$, $\dot{\xi}$ exhibits oscillation around zero with an amplitude of $\sqrt{2(E+K)}$, and both frequencies $\omega_{1,2}$ vary within the interval $\left[\Omega/2 + \sqrt{(E+K)/2}, \Omega/2 - \sqrt{(E+K)/2}\right]$, i.e., two separated frequency intervals merge and become a single interval as $K$ crosses $K^c$. In other words, both frequencies share a phase space domain with an identical invariant measure when $K$ crosses $K^c$.
This analytical analysis of MS has also been studied numerically by employing the Poincar\'e section analysis~\citep{tian13_mplb}. These Poincar\'e sections are constructed in the projected phase planes $(\theta_1, \omega_1)$ and $(\theta_2, \omega_2)$, respectively. This investigation provides convincing evidence supporting the assertion that the occurrence of MS is intrinsically associated with the phenomenon of separatrix crossing, which serves as the overarching mechanism underlying MS.
\section{Types of MS transition}
\label{sec:sec_tran}
The term `MS transition' denotes the transition from a desynchronized state to a MS state, which occurs at a critical value of the coupling parameter. The synchronized state can manifest as either a quasiperiodic or chaotic solution of the full coupled system. It is useful to categorize the transitions from the desynchronized state to the corresponding synchronized state in this paper as (i) quasiperiodic to quasiperiodic MS and (ii) quasiperiodic to chaotic MS.

\subsection{MS transition: quasiperiodic to quasiperiodic dynamics}
\label{sec:qq}
Let us consider a bidirectionally coupled Hamiltonian systems (the $\phi ^4$-system~\citep{wang03}, to be specific), which can be characterized by the following Hamiltonian:
\begin{eqnarray}
	H &=&\frac{p_1^2}{2}+\frac{q_1^4}{4}+\frac{p_2^2}{2}+\frac{q_2^4}{4}+K_Q \left(q_1-q_2\right)^2, \nonumber\\&=&H_1(q_1,p_1)+H_2(q_2,p_2)+H_{\rm c}.\label{eq:phi4_Hamiltonian}
\end{eqnarray}
The system under consideration involves a real non-negative parameter, denoted by $K_Q$, which represents the coupling strength between two one degree-of-freedom subsystems described by $H_i (q_i, p_i)$. The coupling is provided by $H_{\rm c}$. The corresponding equations of motion in canonical form can be expressed as follows:
\begin{subequations}
	\begin{eqnarray}
		\dot{q}_1&=&p_1, \\
		\dot{q}_2&=&p_2, \\
		\dot{p}_1&=&-q_1^3+2K_{Q} \left(q_2-q_1\right), \\
		\dot{p}_2&=&-q_2^3+2K_{Q} \left(q_1-q_2\right).
	\end{eqnarray}\label{eq:EOM_phi4_qq}
\end{subequations}
Upon reaching the MS state, the long-term average of any variable across distinct sites becomes equal. Hence, utilizing averages of specific physical quantities to represent the transition to MS state between various oscillators becomes beneficial. A suitable candidate for such averages is the long-term average of the bare energies of the interacting oscillators:
\begin{equation}
	\label{eq:bare_eng}
	E_i = \frac{1}{T_f}\int_{0}^{T_f} H_i(q_i, p_i) dt,
\end{equation}
where $T_f$ represents the final time up to which the system has evolved. In a synchronized state, the difference in average bare energies ($\Delta E := E_1 -E_2$) should be nearly zero. A desynchronized state between the interacting oscillators can be inferred from a non-zero value of $\Delta E$ when plotted against the coupling strength parameter.

\begin{figure}[h]
	\centering
	\includegraphics[width= 12.2 cm,height= 15 cm, keepaspectratio]{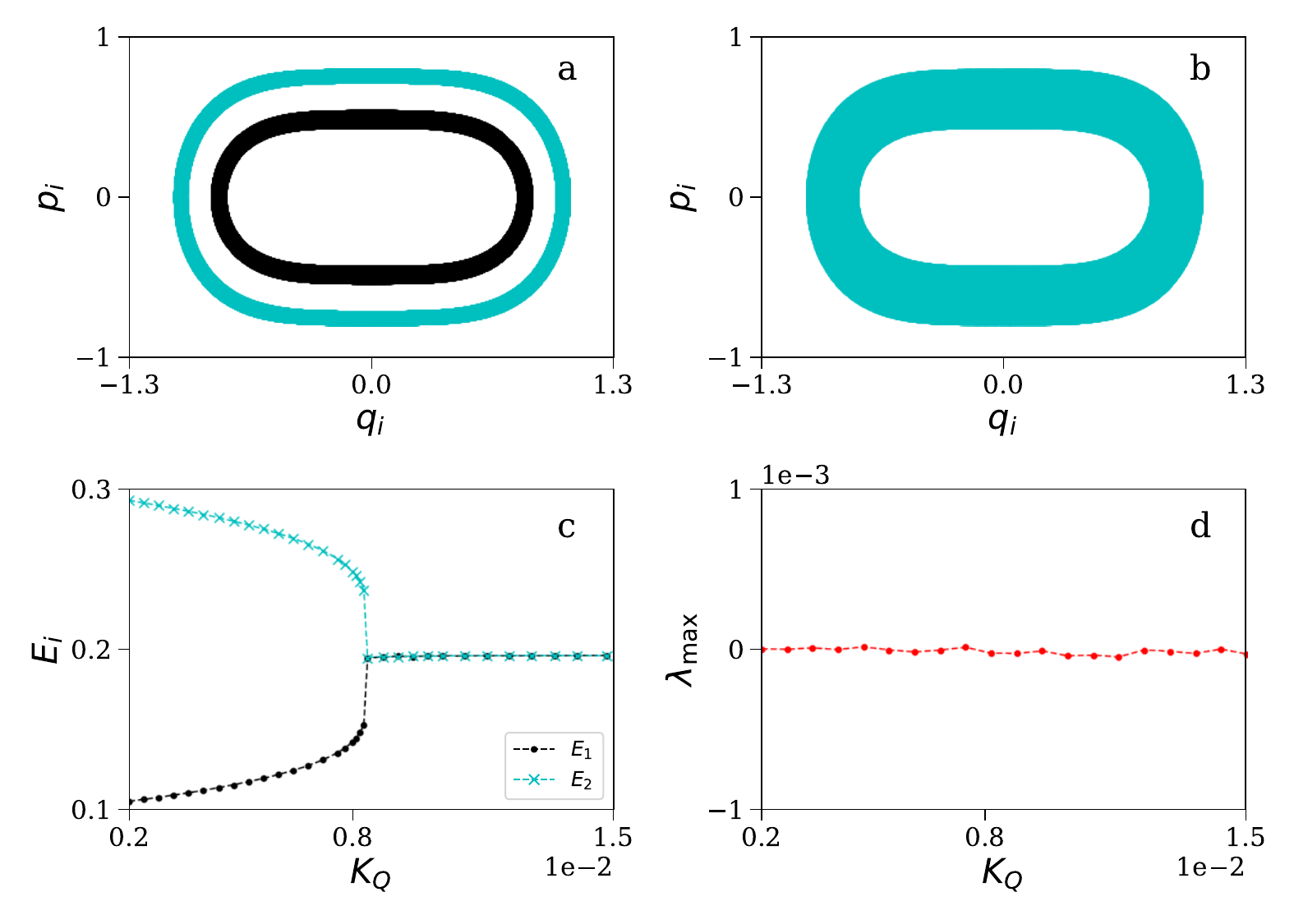}
	\caption{The black and cyan colours in subplots~(a)--(c) correspond to the first and second oscillators. (a) Desynchronized state at $K_Q = 0.6 \times 10^{-2}$. (b) Synchronized state at $K_Q = 1 \times 10^{-2}$. (c) Average bare energies ($E_i$) are calculated, using Eq.~\ref{eq:bare_eng}, as a function of $K_Q$. (d) The maximum Lyapunov exponent ($\lambda_{\rm max}$) is plotted as a function of $K_Q$. In both cases, desynchronized and synchronized, the four-dimensional oscillator (Eq.~\ref{eq:EOM_phi4_qq}) exhibits quasiperiodic oscillations.}
	\label{fig:qq_four}
\end{figure}

The phase space dynamics of a Hamiltonian system is profoundly influenced by the structure of the phase space  and the initial conditions, as is the case with MS, a noteworthy phenomenon. This dependence on initial conditions arises from two primary factors: Firstly, Hamiltonian systems lack attractors, resulting in each orbit displaying distinct asymptotic dynamical behaviour compared to others. Secondly, an initial condition uniquely determines the energy of the autonomous Hamiltonian system, effectively treating the initial condition as a system parameter, a departure from dissipative systems. Consequently, altering the initial conditions inevitably leads to divergent system dynamics; even a slight adjustment may shift dynamics from quasiperiodic to chaotic or vice versa. However, we start investigating the MS transition with initial condition $(q_1(0), q_2(0), p_1(0), p_2(0))$ $ = (0.0, 0.0, \sqrt{0.2}, \sqrt{0.6})$ and $H = 0.4$~\citep{wang02}. After calculating $E_i$ at different values of $K_Q$, we have obtained a transition to MS state while $K_Q $ is varying within the range $[0.2 \times 10^{-2}, 1.5 \times 10^{-2}]$. Figures~\ref{fig:qq_four}a and \ref{fig:qq_four}b depict the projected phase space trajectories in the two-dimensional phase space at $K_Q = 0.6 \times 10^{-2}$ and $1 \times 10^{-2}$, respectively. Figure~\ref{fig:qq_four}a illustrates a desynchronized state showing that the two subsystems occupy distinct regions of the projected two-dimensional phase space. In contrast, we observe the synchronized state at  $K_Q = 1 \times 10^{-2}$ (Fig.~\ref{fig:qq_four}b). The average bare energies $E_i$ are plotted as a function of $K_Q$ in Fig.~\ref{fig:qq_four}c. At lower values of $K_Q$, $E_i$ are non-overlapping; for $K_Q \geq K^c_Q$ (where $ K^c_Q \simeq 0.84\times 10^{-2}$), $E_i$ overlapped, implying the occurrence of MS state in this range of $K_Q$.

Finally, we have calculated the maximum Lyapunov exponent ($\lambda_{\rm max}$) as a function of coupling strength $K_Q$ and plotted in Fig.~\ref{fig:qq_four}d. The maximum Lyapunov exponent~\citep{strogatz07} measures the rate of divergence or convergence of nearby trajectories in phase space. It quantifies the sensitivity of the system's behaviour to small changes in the initial conditions. A positive $\lambda_{\rm max}$ indicates that nearby trajectories in phase space diverge exponentially fast, implying chaotic behaviour. In contrast, a zero $\lambda_{\rm max}$ indicates a quasiperiodic motion. Therefore, Fig.~\ref{fig:qq_four}d supports to conclude that Eq.~\ref{eq:EOM_phi4_qq} exhibits quasiperiodic motion for $K_Q \in [0.2 \times 10^{-2}, 1.5 \times 10^{-2}]$. It is worth noting that the subscript `Q' in $K_Q$ is used to indicate a quasiperiodic system trajectory that is present both before and after the synchronization transition.

\subsubsection{Phase-based order parameters of MS}
Hampton and Zanette~\citep{hampton99} have introduced two order parameters, denoted as $\eta$ and $f$, which serve as tools for discerning the MS transition based on the phase dynamics of the interacting oscillators. Let $\phi_1(t)$ and $\phi_2(t)$ be the instantaneous phases of the respective oscillators calculated using the standard Hilbert transform~\citep{pikovsky01} of phase space variables. In the desynchronized state, the phase difference ($\Delta \phi := \phi_1 - \phi_2$) changes, on average, linearly with evolution time, i.e., $\Delta \phi(t) \approx \eta t + \Delta \phi(0)$. This observation arises due to the difference in main frequencies between the orbits. This order parameter, $\eta$, can be defined as:
\begin{equation}
	\label{eq:eta}
	\eta = \left|\lim\limits_{t \rightarrow \infty}\frac{\Delta \phi(t)}{t}\right|.
\end{equation}
On the contrary, when the two orbits achieve synchronization and share a common domain in phase space, $\Delta \phi(t)$ undergoes oscillations with a characteristic frequency $f$ around a steady value, i.e.,
\begin{equation}
	\label{eq:f}
	\Delta \phi(t) = \Delta \phi(t + 2\pi/f),
\end{equation}
where $f$ is the most dominant Fourier component of $\Delta \phi(t)$. In the synchronized region, $\eta$ reaches to a value of $0$, while $f$ remains finite. Conversely, in the desynchronized region, the opposite scenario emerges. The pronounced changes in the values of $\eta$ and $f$ at this transition point signify its critical nature. However, neither $\eta$ nor $f$ adheres to any power law scaling in the vicinity of the critical point~\citep{hampton99, jing10, gupta17}.

\begin{figure}[h]
	\centering
	\includegraphics[width= 12.0 cm,height= 13 cm, keepaspectratio]{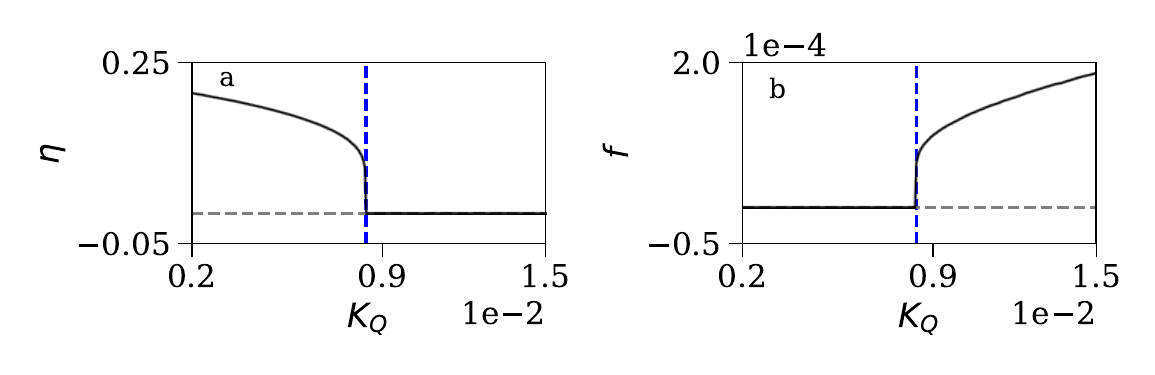}
	\caption{(a) The first phase-based index $\eta$ is plotted as a function of $K_Q$. (b) The second index, $f$, is plotted as a function of $K_Q$. In both subplots, the verticle blue dashed lines correspond to $K_Q = K^c_Q$. The horizontal gray dashed lines correspond to zero magnitudes of these indices.}
	\label{fig:eta}
\end{figure}
These phase-based indices are plotted as a function of $K_Q$ in Fig.~\ref{fig:eta}. The first index $\eta$, in Fig.~\ref{fig:eta}a, has non-zero values initially; on reaching the critical point $K_Q = K^c_Q$, $\eta$ reaches zero and remains zero in the synchronized region ($K_Q > K^c_Q$). The reverse scenario is observed for the second index $f$, as depicted in Fig.~\ref{fig:eta}b.   
\subsubsection{Critical behaviour of MS}
\label{sec:cric_tran}
\begin{figure}[h]
	\centering
	\includegraphics[width= 14 cm,height= 14 cm, keepaspectratio]{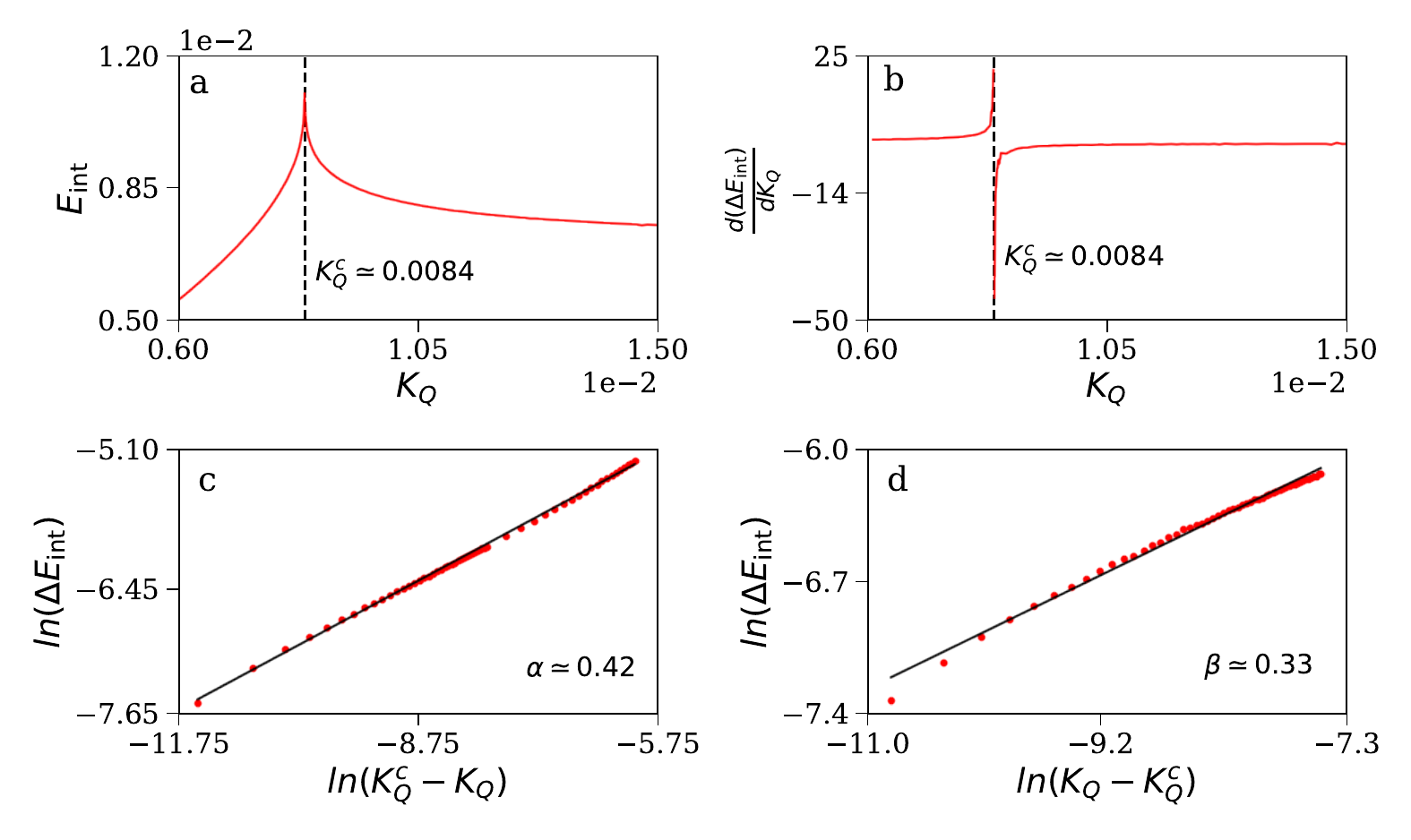}
	\caption{(a) The average interaction energy $E_{\rm int}$ has been plotted as a function of $K_Q$. (b) The first-order derivative, $\frac{d(\Delta E_{\rm int})}{dK_Q}$, is plotted at different values of $K_Q$. Both subplots support the existence of a critical point at $K_Q = K^c_Q$. (c) The critical exponent $\alpha$ has been calculated while Eq.~\ref{eq:EOM_phi4_qq} reaches the critical point. (d) The critical exponent $\beta$ has been calculated after the occurrence of the MS transition.}
	\label{fig:qq_1st_int}
\end{figure}

A critical behaviour of MS has also been examined after calculating another energy-based index, the average interaction energy between the interacting oscillators as function of coupling strength. Along with $\Delta E$, the aforesaid average interaction energy has also been used in the literature to study the transition to MS~\citep{wang03}. We can, therefore, define the average interaction energy ($E_{\rm int}$) between $H_1(q_1,p_1)$ and $H_2(q_2,p_2)$ as follows: 
\begin{equation}
	\label{eq:int_eng}
	E_{\rm int} = \frac{1}{T_f}\left|\int_{0}^{T_f} H_{\rm c} dt\right|,
\end{equation}
where $H_c = K_Q (q_1 - q_2)^2$ for the coupled $\phi^4$-systems (Eq.~\ref{eq:phi4_Hamiltonian}). We have plotted the calculated $E_{\rm int}$ as a function of $K_Q$ in Fig.~\ref{fig:qq_1st_int}a. Initially, $E_{\rm int}$ increases monotonically with the increase in $K_Q$ up to $K_Q = K^c_Q$ and decreases thereafter. In other words, a kink is visible at $K_Q = K^c_Q$. For further confirmation, the first-order derivative of $E_{\rm int}$ with respect to $K_Q$ has been plotted as a function of $K_Q$ (Fig.~\ref{fig:qq_1st_int}b). The discontinuity at $K_Q = K^c_Q$ in Fig.~\ref{fig:qq_1st_int}b mathematically supports the existence of this kink.
In order to study the critical behaviour associated with this MS transition, we can define the difference in interaction energy as $\Delta E_{\rm int} := (E^c_{\rm int} - E_{\rm int})$, where $E^c_{\rm int}$ is the average interaction energy at $K_Q = K^c_Q$. This $\Delta E_{\rm int}$ follows some power laws on either side of critical point $K_Q = K^c_Q$, as mentioned follows:
\begin{equation}
	\Delta E_{\rm int} \propto \begin{cases}
		(K^c_{Q} - K_{Q})^{\alpha}\, \,  \text{for $K_Q < K^c_Q$,}
		\\
		(K_{Q} - K^c_{Q})^{\beta}\, \,  \text{for $K_Q > K^c_Q$},
	\end{cases}
\end{equation}
where $\alpha$ and $\beta$ are the critical exponents. For the example in hand, the calculated critical exponents are $\alpha = 0.42$ and $\beta = 0.33$ (Figs.~\ref{fig:qq_1st_int}c and \ref{fig:qq_1st_int}d).

Wang et al.~\citep{wang03} extensively analyze the scaling law behind MS in coupled $\phi^4$ systems (with a different set of initial conditions) by computing the average interaction energy. They numerically verify different scaling laws before and after MS state, with critical exponents of $1/3$ and $1/2$. We can also detect this critical behaviour of MS in a two-species bosonic Josephson junction~\citep{tian13} and coupled-pendulum system hung from a common beam~\citep{tian19}. In both cases, the power law scaling near the MS transition is associated with a critical exponent of $1/2$. Gupta et al.~\citep{gupta17} report the values of these power law indices as $0.83$ and $1.30$ respectively. In passing, Hampton and Zanette~\citep{hampton99} have also investigated this critical logarithmic singularity; However, due to their use of an averaged quantity as the order parameter for the calculation, they do  not identify the scaling law or the critical exponent.
\subsection{MS transition: quasiperiodic to chaotic dynamics}
\label{sec:qc}

Here, we have adopted another Hamiltonian system, as reported by Ghosh et al.~\citep{ghosh18_2}, to characterize a system exhibiting a quasiperiodic trajectory prior to the MS transition, which subsequently becomes chaotic. The explicit form of Hamiltonian is as follows:
\begin{eqnarray}\label{eq:new_hamiltonian}
	H&=&\frac{I_1^2}{2}+\frac{I_2^2}{2}-\frac{K_{\rm C}}{2} \left[\cos \left(\theta _1-3\theta _2\right) +\cos \left(3\theta _1-\theta _2\right)\right],\nonumber\\
	&=&H_1(I_1)+H_2(I_2)+H_{\rm c}.\quad\,
\end{eqnarray}
The function $H_{\rm c}$ used in Eq.~\ref{eq:new_hamiltonian} differs from the one used in the preceding Hamiltonian (Eq.~\ref{eq:phi4_Hamiltonian}). The coupling strength parameter, $K_{\rm C}$, is a non-negative real number. The corresponding canonical equations of motion are given by:
\begin{subequations}
	\begin{eqnarray}
		\dot{\theta}_1&=&I_1, \\
		\dot{\theta}_2&=&I_2, \\
		\dot{I}_1 &=&-\frac{K_{\rm C}}{2}\left[\sin \left(\theta _1-3\theta _2\right)+3\sin \left(3\theta _1-\theta _2\right)\right], \\
		\dot{I}_2 &=& \frac{K_{\rm C}}{2}\left[3\sin \left(\theta _1-3\theta _2\right)+\sin \left(3\theta _1-\theta _2\right)\right].
	\end{eqnarray}\label{eq:EOM_example}
\end{subequations}
\begin{figure*}[h]
	\centering
	\includegraphics[width= 16.6 cm,height= 18 cm, keepaspectratio]{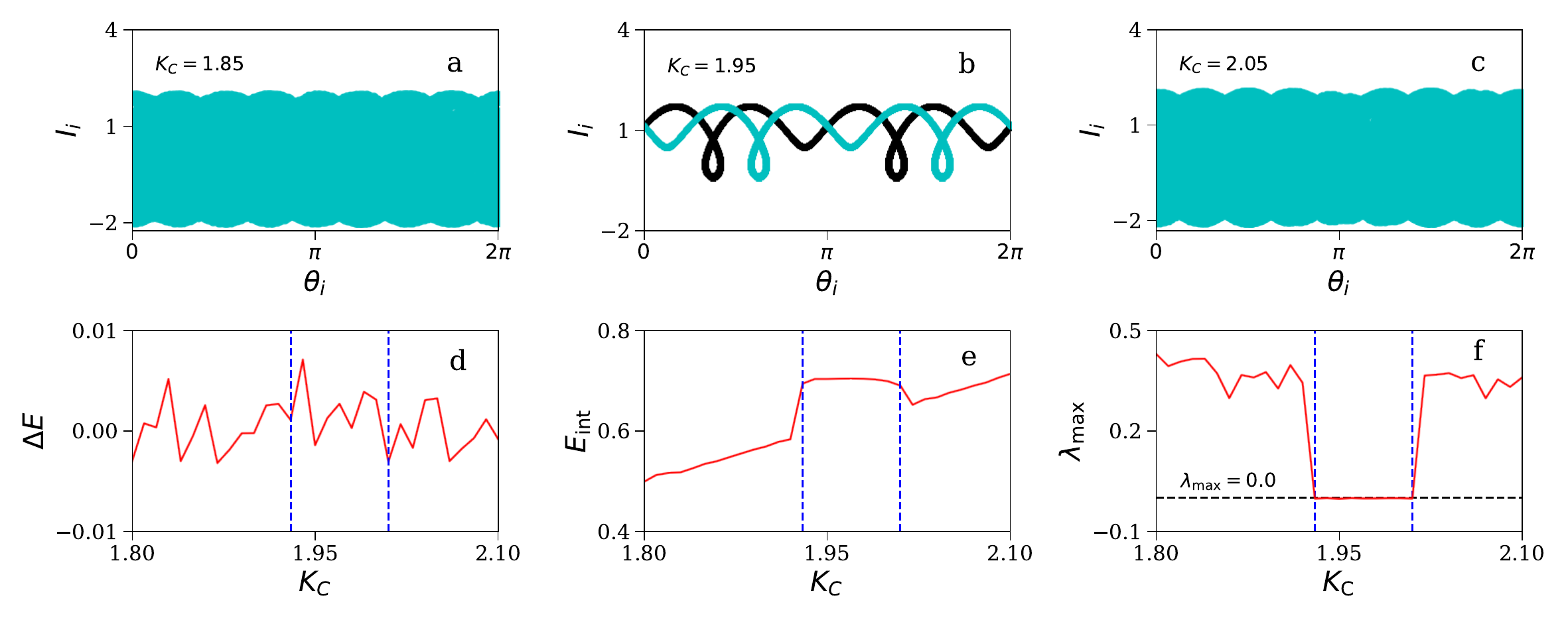}
	\caption{Synchronized state is observed in subplots~(a) and (c) correspond to $K_C = 1.85$ and $2.05$. Subplot~(b) depicts the desynchronized state at $K_C = 1.95$. The black and cyan colours in both plots correspond to the first and second oscillators. (d) The difference in average bare energies ($\Delta E$) is plotted as function of $K_C$. (e) The average interaction energy ($E_{\rm int}$) is plotted as a function of $K_C$. (f) The maximum Laypunov exponent ($\lambda_{\rm max}$) is plotted at different values of $K_C$. The regions within the vertical blue dashed lines in subplots~(d), (e) and (f) represent the desynchronization window.}
	\label{fig:qc}
\end{figure*}
The subscript `C' in $K_{\rm C}$ indicates that the four-dimensional oscillator (Eq.~\ref{eq:EOM_example}) displays chaotic dynamics upon transitioning to the synchronized state. However, similar to the previous example, quasiperiodic dynamics exist in the desynchronized state. When the initial condition is  $(\theta _1(0), \theta _2(0), I_1(0), I_2(0))$ $ = (4.39679, \pi/2, 0.975717, 1.58675)$ with $H = 0.2$, a `desynchronization window' is observed for values of $K_C$ between $1.93$ and $2.01$~\citep{ghosh18_2}. At $K_C = 1.95$, a value within the aforesaid window, the desynchronized state is confirmed by the non-overlapping phase space plots shown in Fig.~\ref{fig:qc}b for the two subsystems. Conversely, Figs.~\ref{fig:qc}a and \ref{fig:qc}c depict synchronized state at $K_C = 1.85$ and $2.05$, respectively. Unlike the previous example, the first index $\Delta E$ fails to detect the desynchronization window (Fig.~\ref{fig:qc}d), while the other index $E_{\rm int}$ effectively identifies it (Fig.~\ref{fig:qc}e). The persistence of long-term chaotic transients contributes to the emergence of non-zero values of $\Delta E$ in the synchronized state~\citep{gupta17}. The convergence of $\Delta E$ towards zero necessitates the evolution of the system for a substantially extended period (i.e., $T_f$ must be very large).
Finally, we have calculated the maximum Lyapunov exponent ($\lambda_{\rm max}$) as a function of $K_C$ and plotted it in Fig.~\ref{fig:qc}f. It ($\lambda_{\rm max}$) has non-zero values outside the desynchronization window and has zero values for $K_C \in [1.93, 2.01]$. Hence, a direct association exists between the MS transition and the transition from quasiperiodicity to chaotic dynamics, and $\lambda_{\rm max}$ can be used as a control parameter to study this kind of MS transition~\citep{wang03, shaoying04}. In other words, if the coupled Hamiltonian system exhibits chaotic dynamics, the interacting oscillators are in MS state. A key feature of chaotic dynamics is their recurrence property, which implies that a chaotic system revisits a specific region of its phase space repeatedly, with arbitrarily small distances between these visits. This key feature is responsible for the aforementioned direct connection.     
Note that the indices $\eta$ (Eq.~\ref{eq:eta}) and $f$ (Eq.~\ref{eq:f}) are not considered as recommended order parameters to study MS transition from quasiperiodic to chaotic dynamics. While $\eta$ can identify quasiperiodic desynchronized regions, it can not identify the MS transition from quasiperiodicity to chaos. This limitation originates from the fact that oscillators are not precisely phase-locked in the chaotic region, and the phase difference ($\Delta \phi$) exhibits arbitrary jumps of $n\pi$ radians between synchronous segments. Furthermore, within the chaotic regime, there might not exist a dominant Fourier component, referred to as $f$, of the dynamical variable $\Delta \phi$. Consequently, the Fourier component $f$ becomes nonexistent in the synchronized region. Besides, the critical transition, as discussed in Sec.~\ref{sec:cric_tran}, does not visible in this category of MS transition~\citep{wang03}.
\subsection{Choice of initial condition}
\label{sec:ini_condi}
\begin{figure}[h]
	\centering
	\includegraphics[width= 10 cm,height= 12 cm, keepaspectratio]{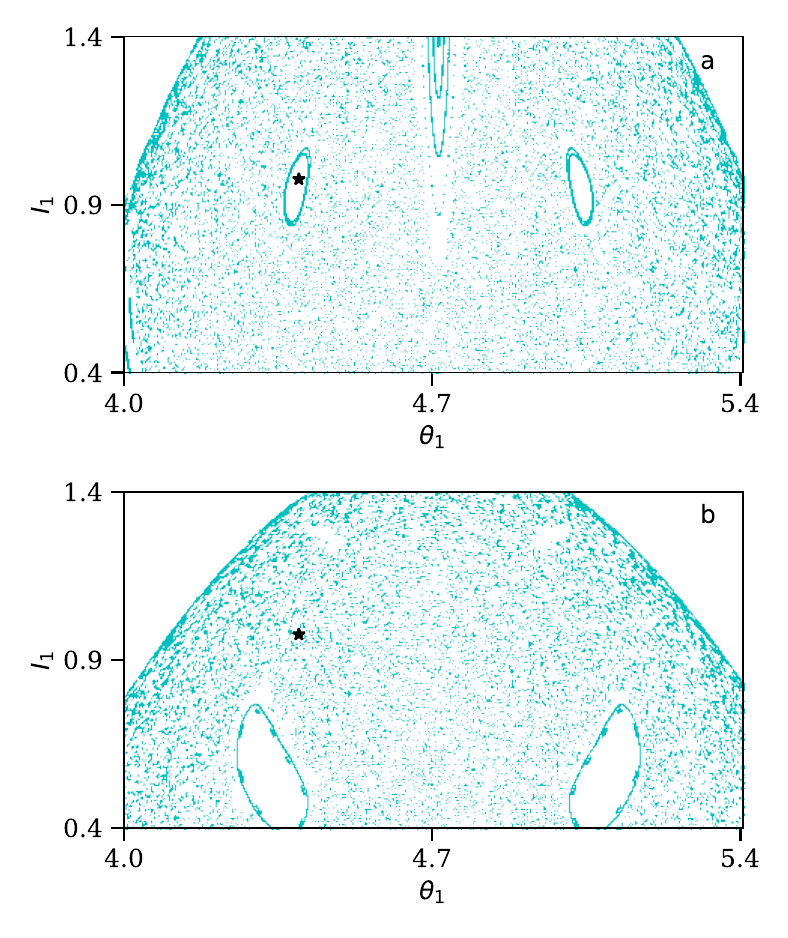}
	\caption{The Poincar\'e sections of the first subsystem are presented in subplots (a) and (b), utilizing fifty distinct sets of initial conditions. Subplot (a) corresponds to $K_{\rm C} = 1.95$, while subplot (b) is associated with $K_{\rm C} = 0.975$. Notably, the black stars in both subplots represent the initial condition, $(\theta _1(0), I_1(0))$  $ = (4.39679, 0.975717)$, used in Sec.~\ref{sec:qc}. A desynchronized state is observed if an initial condition lies within the resonance islands; conversely, if it falls outside these regions, an MS state manifests.}
	\label{fig:ini_condi}
\end{figure}
The selection of an initial condition plays an important role in a Hamiltonian system~\citep{lichtenberg92}. Consequently, each orbit has the potential to exhibit asymptotically distinct dynamical behaviours in comparison to any other orbits. In a Hamiltonian system, an initial condition prescribes the total energy of the autonomous system. This implies that the initial condition effectively plays the role of a system parameter, in contrast to the behaviour observed in dissipative systems~\citep{strogatz07}. Hence, it becomes apparent that even a slight alteration in the choice of initial conditions can induce a considerable transformation in the system's behaviour. A mere modification in the initial condition can instigate a shift in dynamics, potentially transitioning from chaotic to quasiperiodic behaviour or vice versa.
Here, we adopt the Poincar\'e section analysis~\citep{tian13_mplb} to study the effect of choosing an appropriate initial condition in Hamiltonian systems. In order to employ the Poincar\'e section analysis, we evolve Eq.~\ref{eq:EOM_example} for a fixed value of $K_{\rm C}$ and collect $\left(\theta _1(t), I_1(t)\right)$ values when $\theta _2(t) = \pi/2$ and $I_2(t) > 0$. Finally, we repeat this process for different sets of initial conditions with the constraint that the total Hamiltonian value remains unaltered (i.e., $H = 0.2$ in Eq.~\ref{eq:new_hamiltonian}). Figure~\ref{fig:ini_condi} depicts the existence of a few resonance islands in the chaotic sea for two different values of $K_{\rm C}$. As depicted in Fig.~\ref{fig:ini_condi}a, the system characterized by a constant value of coupling strength $K_{\rm C} = 1.95$, there exists a collection of initial conditions situated within the resonance islands yield quasiperiodic motion. The specific initial condition --- $(\theta _1(0), I_1(0))$  $ = (4.39679, 0.975717)$, which eventually emerges the desynchronized state --- adopted in Sec.~\ref{sec:qc} lies within one such resonance island, as shown using the black star marker in Fig.~\ref{fig:ini_condi}a. In the other subplot, Fig.~\ref{fig:ini_condi}b, the location of the aforesaid adopted initial condition is no longer within those islands but in the chaotic sea. Consequently, a change in dynamics is detected, which further yields the occurrence of an MS state. Note that the identification of chaotic dynamics is challenging when the initial conditions are close to the outer boundaries of these islands. In such cases, a trajectory that initially displays regular behaviour over an extended time interval and eventually reveals its intrinsic chaotic nature after a long transient. The corresponding trajectory is generally termed  `sticky orbit'~\citep{dubeibe18}. In the case of sticky orbits, one must evolve the system for an immense amount of time to confirm the MS state.
The conclusion of Fig.~\ref{fig:ini_condi} remains invariant even though we analyze the Poincar\'e sections in the $\left(\theta _2, I_2\right)$ plane. Similar resonance islands are also detected while we extend our study to the MS transition of the first category (Sec.~\ref{sec:qq}). Any initial condition outside of these islands leads to the occurrence of an MS state~\citep{ghosh18_2}.
\subsection{Other indices to detect MS}
MS in a two coupled bosonic Josephson junction model has been studied using the Poincar\'e section analysis~\citep{tian13} and reported that the separatrix crossing is responsible to reaching the MS state. In addition to these techniques, one can always return to the first principle of MS and compare the joint probability density functions of the coupled systems to verify whether the systems share unaltered measures quantitatively~\citep{ghosh18_2}.

Another article~\citep{vincent05_sch} demonstrates the presence of MS in a Hamiltonian system associated with the nonlinear Schr\"odinger equation. Furthermore, it discusses the transition from quasiperiodic measure desynchronization to quasiperiodic MS and from quasiperiodic measure desynchronization to chaotic MS are generic characteristics that define the relationship among coupled nonlinear Schr\"odinger equation subsystems. A recent paper~\citep{tian23} investigates MS in a two-population network of coupled metronome systems. The system comprises multiple identical metronomes on two swing boards connected by a spring. Each swing board represents a single population. The primary objective is to examine the collective dynamics of the multi-population Hamiltonian induced by the coupling strength, which includes inter-population and intra-population couplings. MS and partial MS are observed—the occurrence of these MS states is studied by analyzing the Poincar\'e sections of the Hamiltonian system. 

The study of MS has additionally been carried out within a framework comprising of two nonlinearly coupled oscillators~\citep{gupta17,de18}. In these studies, different types of dynamical states, such as quasiperiodic, chaotic, MS, and desynchronized, along with their respective transitions from one to another, have been observed and quantified by varying the coupling parameter. 

Until now, our discussions have revolved around MS in Hamiltonian systems constituted of two interacting subsystems. In Sec.~\ref{sec:pms}, we explore MS in the Hamiltonian systems composed of more than two interacting subsystems. Analogous to the three-body problem in classical mechanics, the coupled three-oscillator system unveils a richer and more intricate dynamic landscape compared to its two-oscillator counterpart. This intricate behaviour introduces novel states and transition phenomena. This entails extending prior notions and research about MS while retaining the essential insights they offer.

\section{Partial MS}
\label{sec:pms}
MS has been a focal point of inquiry in many-body quantum systems, particularly within the domain of quantum spin chains. In quantum communication theory, these quantum spin chains are envisioned as potential `quantum wires' capable of connecting quantum devices. In the literature, significant research efforts have been directed toward examining spin chains as convincing quantum channels necessary for quantum state transfer and entanglement dynamics~\citep{bose03,christandl04,subrahmanyam04}. When two quantum spin chains are coupled, they initiate the redistribution of quantum correlations within each chain and establish correlations between themselves as time evolves. The quantum counterpart of MS is observed in a pair of coupled quantum kicked Harper chains~\citep{harper55}, where coupling transpires between two spin chains through a potential that varies with time and site~\citep{sur20}. Besides, MS has been investigated in a two-species bosonic Josephson junction~\citep{qiu14}, another example of many-body quantum systems. This two-species bosonic Josephson junction can be regarded as a composite of two interlinked single-species bosonic Josephson junctions~\citep{smerzi97}. Many-body quantum problems encompass a broad class of physical inquiries concerning the attributes of microscopic systems composed of numerous interacting particles~\citep{fabrocini02,fetter03}. This domain incorporates a diverse range of problems that hold fundamental significance in fields like chemistry, physics, and materials science~\citep{fabrocini02,fetter03}.

The Fermi-Pasta-Ulam-Tsingou (FPUT) problem is a seminal nonlinear dynamics and statistical mechanics puzzle~\citep{ford92,dauxois08}. Fermi, Pasta, Ulam, and Tsingou conducted numerical simulations on a chain of particles connected by nonlinear springs. They aimed to understand how energy distributes among different vibrational modes in that one-dimensional chain. They anticipated that energy would spread among the modes equitably, a concept known as equipartition. However, they observed a phenomenon termed `recurrence', wherein the system returns remarkably close to its initial state after a certain period of time. This unexpected behaviour challenged established notions about energy dissipation and highlighted the intricate and surprising dynamics that can emerge in many-body nonlinear systems.

Besides, Kuramoto and Battogtokh~\citep{kuramoto02} reported an important phenomenon in $2002$. A population of identical phase oscillators arranged in a ring-like geometry and subjected to a nonlocal coupling. It has been ascertained that the population is divided into two distinct subgroups of oscillators: one displaying synchrony and the other manifesting desynchrony. In other words, a symmetry breaking can transpire and lead to the emergence of desynchronous behaviour from a state of complete synchrony among identical oscillators. This study~\citep{kuramoto02} sparked considerable astonishment within the nonlinear dynamics community. Anyway, an intriguing state is observed where some subsystems achieve coherence while others have certain degrees of incoherency.

Partial synchronization~\citep{scholl21}, also known as cluster~\citep{su20} (or group~\citep{williams13}) synchronization, characterizes a phenomenon in which some, but not all, components within a system exhibit identical behaviour. An illustrative instance of partial synchronization manifests in the form of chimera states~\citep{abrams04}, which represent states of asymmetric coherence and incoherence within the system's dynamics. In practical scenarios, chimera states find relevance in phenomena like the unihemispheric sleep patterns~\citep{rattenborg00} of birds and dolphins, where one hemisphere slumbers while the other remains alert. The adjustment of inter-hemispheric coupling leads to an intermediate state of incoherence in one hemisphere, giving rise to a chimera-like partial synchronization pattern~\citep{scholl21}. Spontaneous occurrences of chimera states have also been noted in populations of interconnected photosensitive chemical oscillators~\citep{nkomo13}. Similarly, investigations on partial synchronization states extend to diverse networked systems, including epileptic seizures~\citep{jirsa14}, power grids~\citep{motter13}, and social systems~\citep{avella14}.

Here, we are interested in studying partial synchronization in coupled Hamiltonian systems. The partial MS~\citep{wang02} can be observed in systems where there are more than two subsystems coupled appropriately, for example, in systems with three subsystems. Specifically, it has been observed that the first subsystem may be in a synchronized state with the third subsystem but not with the second subsystem. This phenomenon is referred to as partial MS~\citep{vincent05,de18}. Wang et al.~\citep{wang02} conduct an investigation on partial MS in the Bambi, Baowen, and Hong model~\citep{hu00}, which serves as a representation of heat conduction in one-dimensional non-integrable systems. A transition from the partial MS state to the complete MS state has been studied for three coupled double-well Duffing Hamiltonian systems~\citep{vincent05}. Employing both frequency-based and wavelet-based analyses, the partial MS has also been scrutinized in the presence of nonlinear coupling in the classical SU(2) Yang-Mills-Higgs Hamiltonian system~\citep{matinyan03}, which has three degrees of freedom~\citep{de18}. In a recent paper~\citep{tian23}, partial MS has been studied in a network consisting of two populations of coupled metronome systems. This system encompasses numerous identical metronomes arranged on two swing boards interconnected by a spring, each representing an individual population.
\begin{figure}[htbp!]
	\centering
	\includegraphics[width= 12 cm,height= 20 cm, keepaspectratio]{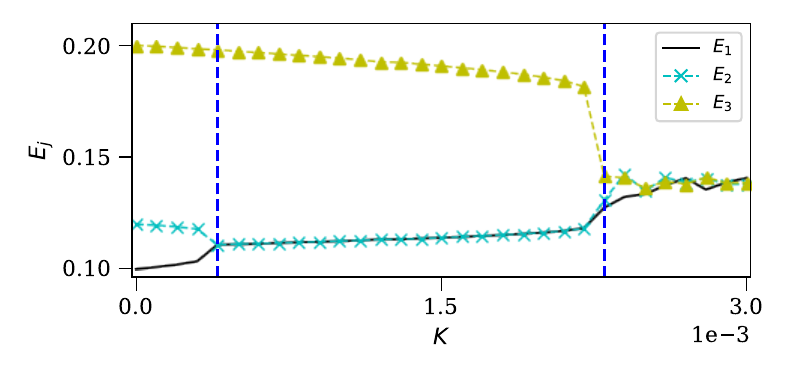}
	\caption{The average bare energies ($E_j$) for Eq.~\ref{eq:EOM_pms} are calculated as a function of $K$ using Eq.~\ref{eq:bare_eng}. Two vertical blue dashed lines correspond to $K = 0.4 \times 10^{-3}$ and $2.3 \times 10^{-3}$.}
	\label{fig:partial_ms_enrg}
\end{figure}

\begin{figure}[htbp!]
	\centering
	\includegraphics[width= 9.5 cm,height= 12 cm, keepaspectratio]{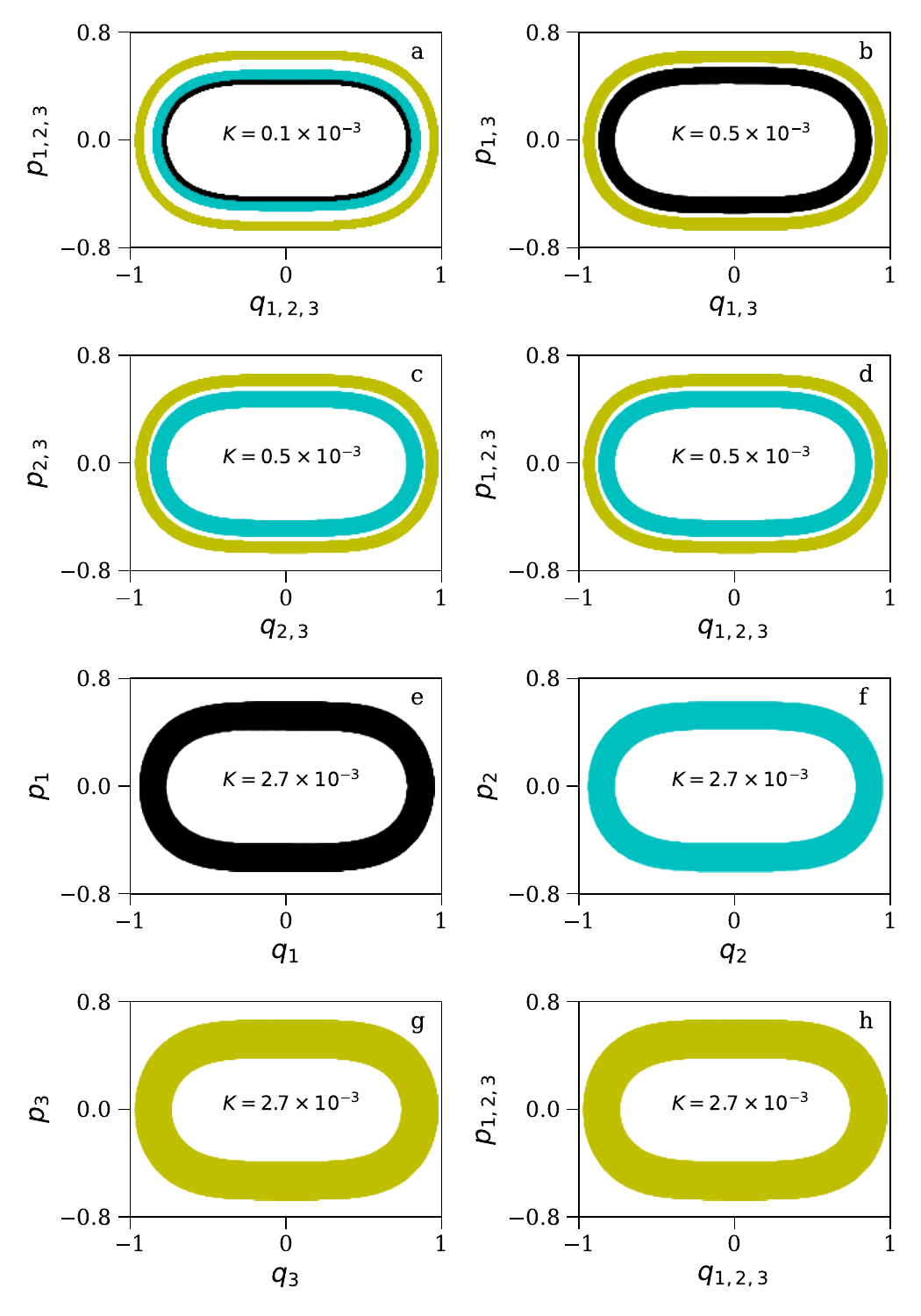}
	\caption{The black, cyan, and yellow colours correspond to the first, second, and third oscillators of Eq.~\ref{eq:EOM_pms}. Subplot (a) depicts that all three oscillators are in the desynchronized state at $K = 0.1 \times 10^{-3}$. Subplots (b)--(d) describe that the first and second oscillators are in MS state at $K = 0.5 \times 10^{-3}$. However, none of them is in MS state with the third one. Subplots (e)--(h) show that all three oscillators are in MS state at $K = 2.7 \times 10^{-3}$.}
	\label{fig:partial_ms_phase}
\end{figure}
This study uses the classical $\phi^4$-system to scrutinize partial MS. In Sec.~\ref{sec:qq}, we have used the $\phi^4$-system to study the MS transition. Our approach encompasses the nearest neighbour diffusive coupling and implements the periodic boundary conditions. The Hamiltonian governing the system is as follows:
\begin{equation}
	H = \frac{p_j^2}{2}+\frac{q_j^4}{4}+ \frac{K}{2} \left(q_{j+1}-q_j\right)^2,\label{eq:pms_Hamiltonian}
\end{equation}
where $j = 1, 2, \cdots, N$. The dynamics of the $j^{\rm th}$ oscillator is precisely determined through the following canonical equations:
\begin{subequations}
	\begin{eqnarray}
		\dot{q}_j&=&p_j, \\
		\dot{p}_j&=&-q_j^3+ K \left(q_{j+1} + q_{j-1} -2q_j\right).
	\end{eqnarray}\label{eq:EOM_pms}
\end{subequations}
The implementation of periodic boundary condition requires $q_{N+1} = q_{1}$ and $q_{0} = q_{N}$. 
\begin{figure}[htbp!]
	\centering
	\includegraphics[width= 9 cm,height= 20 cm, keepaspectratio]{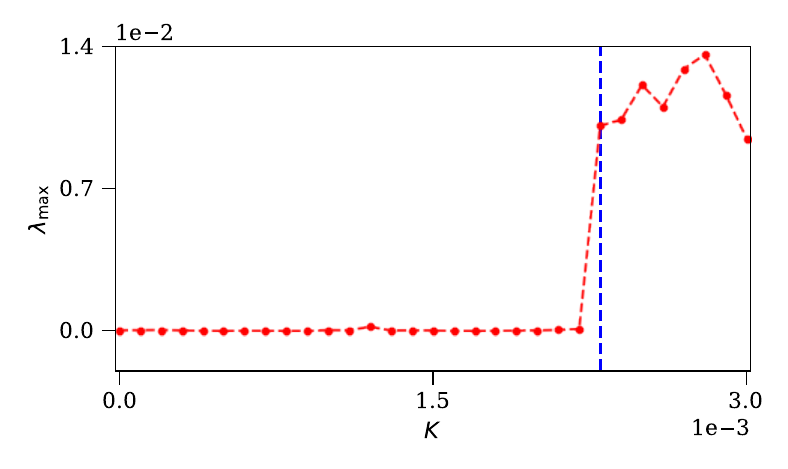}
	\caption{The maximum Lyapunov exponent ($\lambda_{\rm max}$) of Eq.~\ref{eq:EOM_pms} is plotted as a function of $K$. The vertical blue dashed line correspond to $K = 2.3 \times 10^{-3}$. A transition from quasiperiodic to chaotic dynamics is detected at the aforesaid value of $K$.}
	\label{fig:mle_three}
\end{figure}
However, here, we adopt three oscillators (i.e., $N = 3$ in Eq.~\ref{eq:EOM_pms}) and initial condition $(q_1(0), q_2(0), q_3(0), p_1(0), p_2(0), p_3(0))$ $ = (0.0, 0.0, 0.0, \sqrt{0.2}, \sqrt{0.24}, \sqrt{0.4})$~\citep{wang02}. Figure~\ref{fig:partial_ms_enrg} depicts the variation of average bare energies ($E_j$) of Eq.~\ref{eq:EOM_pms} as a function of $K$. These average energies have been calculated using Eq.~\ref{eq:bare_eng}. For small values of $K$ ($K < 0.4 \times 10^{-3}$), all three oscillators exhibit distinct $E_j$ values, indicating a desynchronized state. A notable observation that distinguishes this scenario from the $N = 2$ case is that within the interval $0.4 \times 10^{-3} < K < 2.3 \times 10^{-3}$, we encounter a situation where $E_1 = E_2 \neq E_3$. This suggests that oscillators corresponding to $j = 1$ and $2$ may achieve MS, while oscillator associated with $j = 3$ remains desynchronized. Consequently, due to its dynamic evolution, the system undergoes a state known as partial MS. Upon further increasing $K$ beyond $K = 2.3 \times 10^{-3}$, we observe complete MS among all three oscillators. In order to confirm the MS state, we have plotted the projected phase portraits in two-dimensional phase space at different values of $K$ in Fig.~\ref{fig:partial_ms_phase}. The colours denoting the first, second, and third oscillators of Eq.~\ref{eq:EOM_pms} are as follows: black, cyan, and yellow, respectively. In Fig.~\ref{fig:partial_ms_phase}a, it is illustrated that, at $K = 0.1 \times 10^{-3}$, all three oscillators exist in a desynchronized state. Figures~\ref{fig:partial_ms_phase}b--\ref{fig:partial_ms_phase}d delineate a scenario where the first and second oscillators are concurrently in the MS state at $K = 0.5 \times 10^{-3}$; however, neither of them reaches the MS state concurrently with the third oscillator. More explicitly, Fig.~\ref{fig:partial_ms_phase}b depicts the sharing of non-overlapping phase space regions of the first and third oscillators. A similar sharing of non-overlapping areas between the second and third oscillators is depicted in Fig.~\ref{fig:partial_ms_phase}c. Finally, in Fig.~\ref{fig:partial_ms_phase}d, the cyan plot completely covers the black plot and yields the occurrence of MS state between the first and second oscillators; however, the yellow plot shares non-overlapping regions and yields the desynchronized state of the third oscillator with both first and second oscillators. Figures~\ref{fig:partial_ms_phase}e--\ref{fig:partial_ms_phase}h present a condition wherein all three oscillators simultaneously occupy a MS state at $K = 2.7 \times 10^{-3}$. In conclusion, the first and second oscillators yield the MS state at $K = 0.4 \times 10^{-3}$, while all three oscillators reach the MS state for $K \geq 2.3 \times 10^{-3}$.
\begin{figure}[htbp!]
	\centering
	\includegraphics[width= 10 cm,height= 20 cm, keepaspectratio]{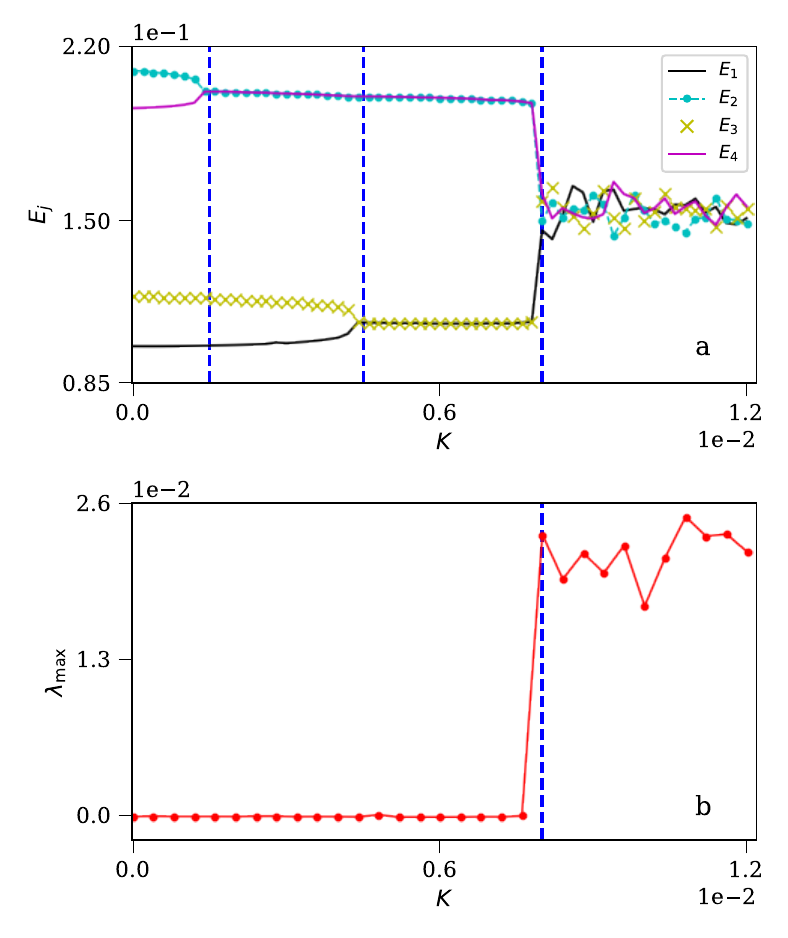}
	\caption{(a) The average bare energies ($E_j$) for four coupled $\phi^4$-systems, following Eq.~\ref{eq:EOM_pms}, are calculated as a function of $K$ using Eq.~\ref{eq:bare_eng}. Three vertical blue dashed lines (from left) correspond to $K = 1.5\times 10^{-3}$, $4.5\times 10^{-3}$ and $8\times 10^{-3}$. (b) The variation of $\lambda_{\rm max}$ as a function of $K$ implies a transition from quasiperiodic to chaotic dynamics is ascertained at $K = 8\times 10^{-3}$. }
	\label{fig:partial_ms_enrg_four}
\end{figure}
This transition from desynchronization to MS is closely linked to a distinct shift in the system's dynamical behaviour. Figure~\ref{fig:mle_three} presents the maximum Lyapunov exponent ($\lambda_{\rm max}$) of the system plotted against $K$. A non-zero value of $\lambda_{\rm max}$ is detected when $K \geq 2.3 \times 10^{-3}$. Therefore, a transition from quasiperiodicity to chaos coincides precisely with the turning point from partial MS to complete MS.
Now, we extend our analysis to four oscillators (i.e., $N = 4$ in Eq.~\ref{eq:EOM_pms}) and adopt the initial condition $(q_1(0), q_2(0), q_3(0), q_4(0), $ $ p_1(0), p_2(0), p_3(0), p_4(0))$ $ = (0.0, 0.0, 0.0, 0.0, \sqrt{0.2}, -\sqrt{0.42}, $  $ -\sqrt{0.24}, \sqrt{0.39})$~\citep{wang02}. We observe four distinct configurations in various coupling strength regions (Fig.~\ref{fig:partial_ms_enrg_four}a). When $K$ is less than $ 1.5\times 10^{-3}$, the average energies of the four oscillators remain distinctly separate, resulting in a desynchronized state. Within the range $1.5\times 10^{-3} < K < 4.5\times 10^{-3}$, a partial MS state manifests. In this state, a pair of oscillators, i.e., oscillators corresponding to $j = 2$ and $4$, exhibit synchronization, while the other two oscillators, $j = 1$ and $3$, remain desynchronized. As $K$ is further increased, within the $4.5\times 10^{-3} < K < 8 \times 10^{-3}$ range, the remaining two oscillators achieve MS, leading to a distinct partial MS state characterized by a higher degree of synchronization. Remarkably, in both regions, $1.5\times 10^{-3} < K < 4.5\times 10^{-3}$ and $4.5\times 10^{-3} < K < 8 \times 10^{-3}$, a noteworthy feature of nonlocal synchronization emerges. This phenomenon is akin to the behaviour observed in dissipative systems~\citep{zhan00}, where two oscillators that are not adjacent in the sequence can exhibit MS. In contrast, the intermediate oscillators remain desynchronized from them. Notably, this measure-symmetry in the nonlocal sites is not inherent in the initial conditions but rather emerges dynamically under specific suitable coupling conditions. All four oscillators attain full MS for sufficiently large coupling values ($K > 8 \times 10^{-3}$). Here, we have $E_1 = E_2 = E_3 = E_4$, which infers that four oscillators constitute a complete MS state.

In Fig.~\ref{fig:partial_ms_enrg_four}b, we observe the relationship between the maximum Lyapunov exponent, $\lambda_{\rm max}$, and the coupling strength $K$. It is clearly visible that $\lambda_{\rm max}$ becomes positive for $K \geq 8\times 10^{-3}$ and remains zero before this threshold. Once more, we observe a consistent pattern where the shift from quasiperiodicity to chaos coincides with the transition from partial MS to complete MS, akin to the observation in Fig.~\ref{fig:mle_three} for the case where $N = 3$. It is noteworthy to emphasize that chaos consistently manifests whenever complete MS occurs among nonlinear oscillators for $N \geq 3$~\citep{wang02,tian19}.
\section{Effects of occasional coupling}
\label{sec:ocs}
In interacting dynamical systems, coupling between subsystems is a fundamental prerequisite for observing synchronization~\citep{pikovsky01}. While this phenomenon was initially discovered in the seventeenth-century~\citep{pikovsky01}, the popularity of synchronization in coupled chaotic systems grew notably after the work by Pecora and Carroll~\citep{pc1990}. Subsequently, one of the most intriguing concepts introduced in this domain is that of `occasional coupling'~\citep{gupte93}. In this framework, the participating subsystems interact intermittently rather than continuously, and synchronization is successfully achieved. It is worth highlighting that, in many instances, occasional coupling yields robust synchronization, even in situations where continuous coupling proves inadequate. In other words, two or more initially desynchronized interacting chaotic systems attain a synchronized state when subjected to occasional coupling pulses~\citep{ghosh20}.

This occasional (or intermittent) interaction can assume either a deterministic or stochastic nature~\citep{ghosh20}. For example, in the stochastic on-off coupling scheme~\citep{jeter15}, the activation or deactivation of the coupling term occurs randomly. In contrast, the interaction follows a deterministic pattern in the on-off coupling scheme~\citep{chen09}. The concept of intermittent coupling has been extended to complex networks incorporating delayed coupling for the examination of synchronization~\citep{sun16}. Employing the occasional coupling, it is possible to render a non-synchronizable network synchronizable~\citep{sch16}. In multi-layered complex networks, investigations on synchronization have incorporated using time-varying inter-layer links~\citep{eser21}. In circadian oscillators within individual cells of fungal systems, stochastic intermittent coupling has been employed to investigate synchronization~\citep{deng16}. Additionally, it has been reported that periodically time-varying changes in coupling among neuron oscillators can enhance synchronization~\citep{parastesh19}. In passing, beyond the purview of synchronization, the occasional coupling is equally effective in reaching the amplitude death in coupled dynamical systems when the continuous coupling fails~\citep{sun18,ghosh22_3}.

The success of an occasional coupling scheme in inducing synchronization in dissipative chaotic systems does not necessarily guarantee its efficacy when employed on previously measure desynchronized systems. This is because occasional coupling in dissipative systems is typically ad hoc, and analytical tools such as conditional Lyapunov exponents~\citep{pecora98} and eigenvalues of the Jacobian matrix of the corresponding linearized transverse dynamics~\citep{ghosh18} are not applicable for characterizing MS. However, the occasional coupling has found application in mitigating measure desynchronization observed in coupled Hamiltonian systems~\citep{ghosh18_2}. More explicitly, the use of the on-off coupling scheme~\citep{chen09} can effectively overcome desynchronization and restore MS between two coupled subsystems~\citep{ghosh18_2}. This study~\citep{ghosh18_2} encompasses both categories of MS states: quasiperiodic and chaotic. This paper undertakes an examination by selecting representative systems from each category and emphasizes the effect of the on-off coupling to retain the MS states in these systems.

In this on-off coupling scheme~\citep{chen09}, the coupling parameter is periodically turned on and off at predetermined intervals. Thus, in the presence of occasional coupling, we need to take care of two time scales: the system time scale ($T_s$) and on-off period ($T$). In order to facilitate a more comprehensive discussion and to provide a clear mathematical representation of the implementation of the on-off coupling scheme~\citep{chen09}, we express the bidirectional coupling between the two Hamiltonian subsystems using the coupling strength $K$ in the following manner~\citep{ghosh18_2}:
\begin{eqnarray}
	\label{vector_field}	
	\dot{\textbf{x}}  &=& \textbf{f}\left(\textbf{x}\right)+ {\chi}(t){K}  {\textsf{C}} \cdot \mathbf{g}\left(\textbf{x}\right),
\end{eqnarray} 
where 
\begin{equation}
	{\chi}(t) = \begin{cases}
		1  \text{ for } nT \le t < (n+\theta)T,\\
		0  \text{ for }(n+\theta)T \le t <(n+1)T,
	\end{cases}
\end{equation}
and $\theta$ is the on-off fraction. By definition, $\theta = 0$ and $1$ refer to the cases when the coupling is inactive, and the continuous coupling is activated between the oscillators, respectively. Hence, the value of $\theta$ must lie between $0$ and $1$, i.e. $0 < \theta < 1$, to keep the occasional coupling activated. Following Eq.~\ref{eq:EOM_phi4_qq} and Eq.~\ref{vector_field}, we have $\textbf{x} = \left(q_1,q_2,p_1,p_2\right)$, $\textbf{f}(\textbf{x}) = (p_1,p_2,-q_1^3,-q_2^3)$, and $\textbf{g}(\textbf{x}) = (0,0,2 q_2 - 2 q_1, -2 q_2 + 2 q_1)$. In addition, $K=K_{\rm Q}$ and the coupling matrix $\textsf{C}_{mn}=\delta_{m3}\delta_{n3}+\delta_{m4}\delta_{n4}$, where $\delta$ is the Kronecker delta. Similarly, one can easily connect Eq.~\ref{eq:EOM_example} with Eq.~\ref{vector_field} and define associated column vectors $\textbf{x}$, $\textbf{f}(\textbf{x})$, and $\textbf{g}(\textbf{x})$ and the coupling matrix $\textsf{C}$.

The solution of Eq.~\ref{vector_field} can be written as:
\begin{eqnarray}
	\label{vector_field_TE}
	\textbf{x}\left(t+T\right)&=&\textbf{x}(t)+\int _t^{t+T}\textbf{f}\left(\textbf{x}(t')\right) dt' +\int _t^{t+T} {\chi}(t')K\textsf{C} \cdot \mathbf{g}(\textbf{x}(t')) dt'.
\end{eqnarray}
If $T$ is significantly smaller than the system time-scale ($T_s$), the functions $\textbf{f}$ and $\textbf{g}$ do not vary much and can be considered constant over the time $T$. Therefore, in the presence of the assumption that $\textbf{f}$ and $\textbf{g}$ remain constant throughout the time interval $T$, we can approximately write $\textbf{x}\left(t+T\right)$ as follows:
\begin{equation}
	\label{approx_vector_field_TE}
	\textbf{x}\left(t+T\right) \approx \textbf{x}(t)+\textbf{f}\left(\textbf{x}(t)\right) T +\theta K\textsf{C} \cdot \mathbf{g}(\textbf{x}(t))T.
\end{equation}
In the last term, we have explicitly integrated the concept that the coupling operates solely during a fraction $\theta$ of the total time $T$. Consequently, it becomes evident that we can conceptualize the system subject to the on-off coupling as equivalent to the system subjected to continuous coupling, albeit with an effectively reduced coupling strength value, denoted as follows:
\begin{equation}
	\label{eq:eff}
	K_{\text{eff}}=\theta K.
\end{equation}
Therefore, for the condition $T \ll T_s$, a linear stability analysis has been performed for coupled oscillators and analytically understand the shifting of the desynchronized window along the higher range of coupling strength~\citep{ghosh18_2}. Although this approximate connection of the on-off coupling and the continuous coupling schemes is inapplicable for $T \sim T_s$, the on-off coupling scheme can lead to MS for $T \sim T_s$~\citep{ghosh18_2}. Besides, the effectiveness of the transient uncoupling scheme, another example of the occasional coupling scheme, has also been reported in the literature to attain MS when the continuous coupling fails~\citep{ghosh18_2}. A general understanding of why and how these occasional coupling schemes work is still missing in the literature. 
In a subsequent study~\citep{tian20}, the occasional coupling has been used to investigate MS in a two-species bosonic Josephson junction. This study reveals that the broken symmetry of nonlocal MS states appearing in the $0$-phase mode and $\pi$-phase mode can be restored by employing the on-off coupling scheme. It has also demonstrated that the nonlocal MS states can be transformed into conventional MS states, either quasiperiodic or chaotic MS states. More specifically, for the $0$-phase mode, the broken symmetry restores by converting the nonlocal MS state into a conventional quasiperiodic MS state. However, for the $\pi$-phase mode, the broken symmetry is restored, and chaotic MS states emerge.
Until now, we have kept ourselves restrictive in the classical regime. MS in quantum systems has also been the subject of extensive research over the last decade, focusing on understanding its fundamental properties and exploring its potential applications in areas such as quantum information processing and quantum communication.
\section{MS in quantum systems}
\label{sec:quantum}
The underlying principle of the MS is rooted in the notion of the area enclosed by the trajectory in phase space. However, this concept of phase space is not applicable in quantum mechanics. Consequently, one cannot apply the concept of the MS directly to quantum systems, but its equivalent can be observed. This section studies the transition to MS in a two-species bosonic Josephson junction (BJJ). From an experimental perspective, the superconducting Josephson junction has emerged as one of the extensively investigated systems in the exploration of synchronization. This specific junction stands as a prominent model of coupled dynamical systems. Moreover, recent advancements in the experimental manipulation of Bose-Einstein condensation have facilitated the creation and control of a BJJ~\citep{albiez05}. A two-species BJJ can be achieved experimentally by confinement of a binary mixture of Bose-Einstein condensations within a symmetric double-well potential. In a seminal theoretical investigation, Smerzi et al.~\citep{smerzi97} have established a mapping between a single-species BJJ and a classical pendulum system. In this section, MS is studied in a two-species BJJ based on semiclassical~\citep{tian13} and quantum~\citep{qiu14} approaches in Sec.~\ref{sec:bjj} and Sec.~\ref{sec:quan}, respectively. 
\subsection{MS using semiclassical approach}
\label{sec:bjj}
Tian et al.~\citep{tian13} have studied MS in quantum systems using a semiclassical approach under the assumption of sufficiently weak interatomic interactions and employing the established two-mode approximation~\citep{leggett01}. The corresponding Hamiltonian in the second quantization is expressed as follows:
\begin{eqnarray}
	\label{eq:EOM_bjj1}
	\hat{H}&=&\frac{u_1}{2N_1}\left[(\hat{a}^{\dagger}_{L} \hat{a}_{L})^2 + (\hat{a}^{\dagger}_{R} \hat{a}_{R})^2 \right] + \frac{u_2}{2N_2}\left[(\hat{b}^{\dagger}_{L} \hat{b}_{L})^2 + (\hat{b}^{\dagger}_{R} \hat{b}_{R})^2\right] -\frac{v_1}{2}\left(\hat{a}^{\dagger}_{L} \hat{a}_{R} + \hat{a}^{\dagger}_{R} \hat{a}_{L}\right) -\frac{v_2}{2}\left(\hat{b}^{\dagger}_{L} \hat{b}_{R} + \hat{b}^{\dagger}_{R} \hat{b}_{L}\right) \nonumber \\
	&& + \frac{u_{12}}{\sqrt{N_1 N_2}} \left(\hat{a}^{\dagger}_{L} \hat{a}_{L} \hat{b}^{\dagger}_{L} \hat{b}_{L} + \hat{a}^{\dagger}_{R} \hat{a}_{R} \hat{b}^{\dagger}_{R} \hat{b}_{R}\right),
\end{eqnarray}
where $\hat{a}_{R}$ and $\hat{a}_{L}$ are the annihilation operators correspond to the localized modes in the right and left wells, respectively, of the first species $a$ with the total number of particles $N_1$. Similarly, $\hat{a}^{\dagger}_{R}$ and $\hat{a}^{\dagger}_{L}$ are the respective creation operators corresponding to right and left wells of the first species. The second species $b$, consists of total $N_2$ particles, has annihilation (creation) operators $\hat{b}_{R}$ ($\hat{b}^{\dagger}_{R}$) and $\hat{b}_{L}$ ($\hat{b}^{\dagger}_{L}$) for the right and left wells, respectively. The parameters $u_i$ (where $i=1,2$) and $u_{12}$ represent the effective interaction strengths associated with atomic collisions involving species of the same kind and species of different kinds, respectively. Finally, $v_i$, the effective Rabi frequency, measures the interaction strength between two wells.

In the semiclassical regime~\citep{smerzi97,raghavan99,leggett01}, the evolution of the system can be characterized through a classical Hamiltonian denoted as $H = \left\langle \Psi| \hat{H} | \Psi \right\rangle$, where $| \Psi \rangle = \frac{1}{\sqrt{N_1}} (\alpha_L \hat{a}^{\dagger}_L + \alpha_R \hat{a}^{\dagger}_R)^{N_1}|0,0\rangle \otimes \frac{1}{\sqrt{N_2}} (\beta_L \hat{b}^{\dagger}_L + \beta_R \hat{b}^{\dagger}_R)^{N_2}|0,0\rangle $ is a collective state with total number of particles $N = N_1 + N_2$. In this context, four c numbers, $\alpha_L = |\alpha_L| e^{i \theta_{1L}}$, $\alpha_R = |\alpha_R| e^{i \theta_{1R}}$, $\beta_L  = |\beta_L| e^{i \theta_{2L}}$, and $\beta_R  = |\beta_R| e^{i \theta_{2R}}$, represent the probability amplitudes associated with the presence of two distinct species of atoms within the two wells. The conservation of particle numbers for each species mandates that $|\alpha_L|^2 + |\alpha_R|^2 = 1$ and $|\beta_L|^2 + |\beta_R|^2 = 1$. Upon the introduction of the relative population differences, denoted as $S_1 = (|\alpha_L|^2 - |\alpha_R|^2)$ and $S_2 = (|\beta_L|^2 - |\beta_R|^2)$, along with the relative phase differences $\theta_1 = (\theta_{1L} - \theta_{1R})$ and $\theta_2 = (\theta_{2L} - \theta_{2R})$, we derive the mean-field Hamiltonian as presented below:
\begin{equation}
	H = H_1(\theta_1,S_1)+H_2(\theta_2,S_2)+H_{\rm c},\label{eq:bbj_Hamiltonian}
\end{equation}
where 
\begin{equation}
	H_i(\theta_i,S_i) = -\frac{u_i}{2} S_i^2 + v_i \sqrt{1-S_i^2}\cos \theta_i,
\end{equation}
and 
\begin{equation}
	H_c = -u_{12} S_1 S_2.
\end{equation}
Therefore, a two-species BJJ exhibits similarities to two interlinked single-species BJJs. It is evident that the coupling arises due to the interspecies interaction $u_{12}$. The equations of motion can be deduced as follows:
\begin{subequations}
	\begin{eqnarray}
		\dot{\theta}_1&=& - u_1 S_1 - \frac{v_1 S_1}{\sqrt{1 - S_1^2}} \cos \theta_1 - u_{12} S_2, \\
		\dot{S}_1&=& v_1 \sqrt{1 - S_1^2} \sin \theta_1, \\
		\dot{\theta}_2&=& - u_2 S_2 - \frac{v_2 S_2}{\sqrt{1 - S_2^2}} \cos \theta_2 - u_{12} S_1, \\
		\dot{S}_2&=& v_2 \sqrt{1 - S_2^2} \sin \theta_1.
	\end{eqnarray}\label{eq:EOM_bjj}
\end{subequations}

\begin{figure}[htbp!]
	\centering
	\includegraphics[width= 10.5 cm,height= 14 cm, keepaspectratio]{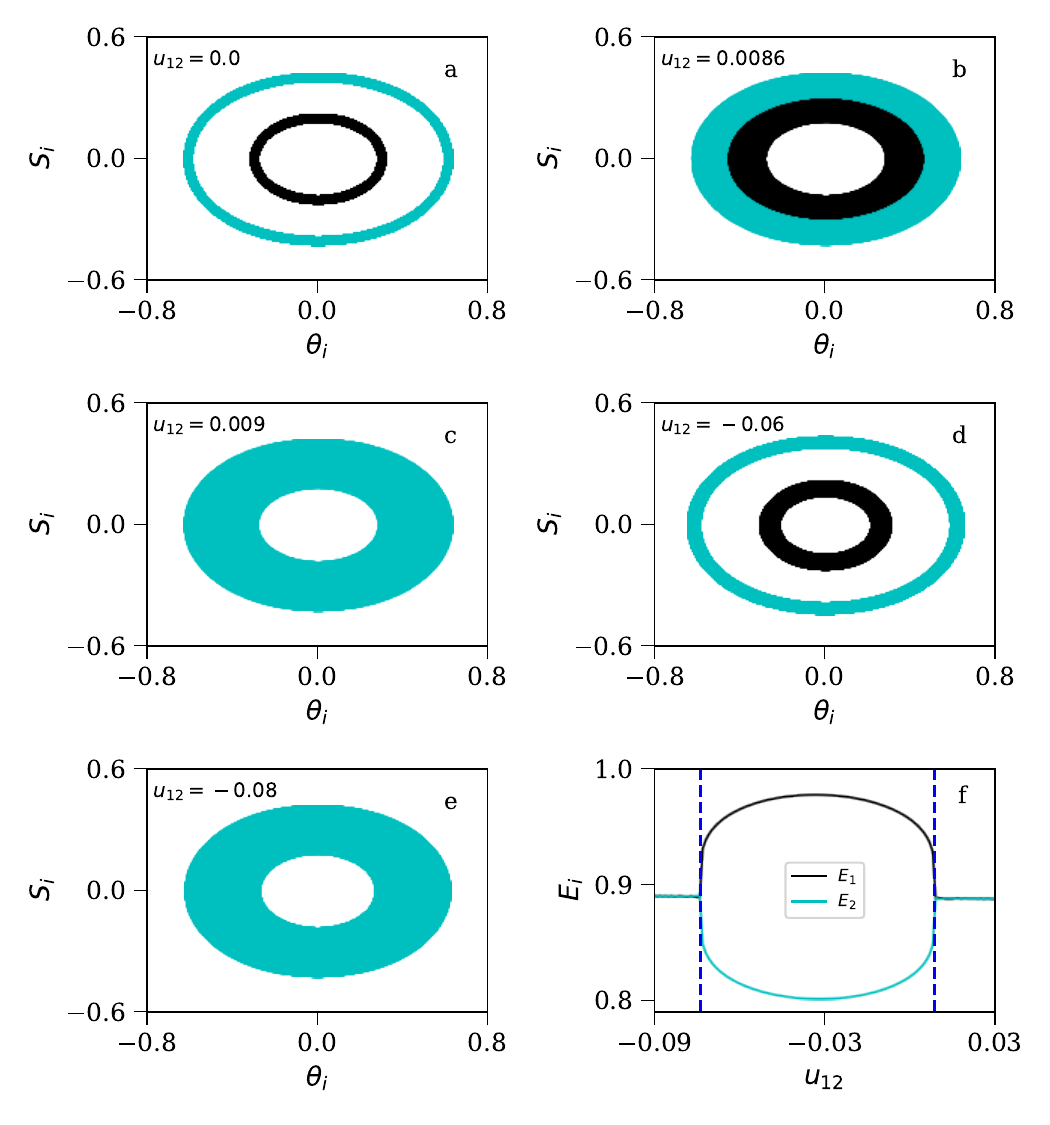}
	\caption{The black and cyan colours correspond to the first and second oscillators of Eq.~\ref{eq:EOM_bjj}. Subplot~(a) represents phase space diagram when no interspecies interaction is activated. Transition to MS is studied in the presence of repulsive (subplots~(b) and (c)) and attractive (subplots~(d) and (e)) interspecies interactions. The average energies of each species are plotted as a function of the interspecies interaction $u_{12}$ in subplot~(f). The vertical blue dashed lines correspond to $-7.38 \times 10^{-3}$ and $u_{12} = 8.6 \times 10^{-4}$. } 
	\label{fig:qms_bjj}
\end{figure}

As our focus lies in illustrating the influence of coupling on the dynamics of each species, we represent the collective motion by projecting the state of the complete system onto the respective individual phase spaces. We adopt the parameter values $u_1 = u_2 = 1.2$ and $v_1 = v_2 = 1$, and the initial condition $(\theta_1(0), S_1(0)), \theta_2(0), S_2(0) =$ $ 
(0.0, 0.2, 0.0, 0.4)$~\citep{tian13}. To this end, following Eq.~\ref{eq:bare_eng}, we can define the average energy of each single-species BJJ as follows: 
\begin{equation}
	\label{eq:bare_eng_bjj}
	E_i = \frac{1}{T_f}\int_{0}^{T_f} H_i(\theta_i, S_i) dt.
\end{equation}
Figures~\ref{fig:qms_bjj}b and \ref{fig:qms_bjj}c depict the temporal evolution of MS under the influence of repulsive interspecies interactions (i.e., $u_{12} > 0$). As the coupling strength $u_{12}$ is monotonically increased, we trace the trajectories on the $(\theta_i, S_i)$ phase plane for the two subsystems. For $u_{12} = 0$, as illustrated in Fig.~\ref{fig:qms_bjj}a, the initial conditions correspond to two distinct quasiperiodic orbits, forming enclosed curves. However, as $u_{12}$ is raised above zero, these two enclosed curves transform into smooth quasiperiodic trajectories that meander within separate phase-space regions, adopting a ring-like shape. With the continued increase in $u_{12}$, the two phase-space regions undergo a gradual transformation, wherein the outer boundary of the inner domain approaches the inner boundary of the outer domain. They draw nearer to each other until $u_{12}$ reaches a critical value of $8.6\times 10^{-4}$, as depicted in Fig.~\ref{fig:qms_bjj}b, at which point the two converging boundaries nearly touch. Beyond this threshold $u_{12} = 8.6\times 10^{-4}$, the MS state between these two species $a$ and $b$ is detected, as exemplified in Fig.~\ref{fig:qms_bjj}c. The previously distinct phase-space domains, formally well-separated, converge and envelop the phase-space regions with indistinguishable invariant measures.

Figures~\ref{fig:qms_bjj}d and \ref{fig:qms_bjj}e portray the progression with increasing strength of attractive interspecies interactions. We again start with Fig.~\ref{fig:qms_bjj}a when no coupling is activated. For values of $u_{12}$ less than zero (i.e., $u_{12} < 0$), we can also reach MS from the initial desynchronized state. Non-overlapping ring-like bands are detected at $u_{12} = -0.06$ (Fig.~\ref{fig:qms_bjj}d). Finally, MS has been detected at $u_{12} = -0.08$ (Fig.~\ref{fig:qms_bjj}e). Note that routes to MS are different for attractive and repulsive interspecies interactions~\citep{tian13}. In Fig.~\ref{fig:qms_bjj}f, we present the average energies $E_1$ and $E_2$ as functions of the interspecies interactions denoted by $u_{12}$. Notably, distinct sharp transitions occur at  $u_{12} = -7.38 \times 10^{-3}$ and $u_{12} = 8.6 \times 10^{-4}$ for the cases of repulsive and attractive interactions, respectively. These two transition points are shown using two verticle blue dashed lines in Fig.~\ref{fig:qms_bjj}f. A finite disparity between $E_1$ and $E_2$ is evident when the value of $u_{12}$ lies within the region bounded by the blue verticle dashed lines. Conversely, both species exhibit identical average energies when the value of $u_{12}$ remains outside the aforesaid boundary. 
The semiclassical theory has revealed the existence of two distinct dynamic regimes~\citep{smerzi97, raghavan99, fu06}  of a single-species BJJ: the Josephson oscillation regime and the self-trapping regime. In our discussion, we have examined MS in the Josephson oscillation regime, where the variable $\theta_i$ oscillates in proximity to $\theta_i = 0$, signifying the zero-phase mode. However, it is noteworthy that MS has also been observed in the self-trapping regime~\citep{tian13}, where $\theta_i$ undergoes oscillations around $\theta_i = \pi$, corresponding to the $\pi$-phase mode. Six different scenarios of MS, including two in the $0$-phase mode (Fig.~\ref{fig:qms_bjj}) and four in the localized and nonlocalized $\pi$-phase modes, are characterized, and common features behind them are revealed. The MS transition corresponds to the sudden merger of the average energies of the two species. The power law scaling with a critical exponent of $1/2$ is verified for all scenarios. Moreover, a three-dimensional view of MS is provided, revealing features that are not visible in the two-dimensional phase space. Poincar\'e section analysis shows that a two-species bosonic Josephson junction exhibits separatrix crossing behaviour at the critical coupling intensity. It is concluded that separatrix crossing is the general mechanism underlying the different scenarios of MS transitions in the two-species bosonic Josephson junction. In the case of MS transition from quasiperiodic to quasiperiodic dynamics, the maximum Lyapunov exponent ($\lambda_{\rm max}$) of the Hamiltonian system is positive at the transition point (the point of separatrix crossing)~\citep{tian13_mplb}. However, this point is singular and unmeasurable~\citep{tian13_mplb}.
\subsection{MS using quantum approach}
\label{sec:quan}
A close analogy of MS in quantum scenario has been made by Qiu et al~\citep{qiu14}. They have characterized the quantum many-body MS in a three-dimensional (3D) space defined by the average value of the pseudoangular momentum. The dynamics of two quantum many-body systems become coupled through particle-particle interactions. The Hamiltonian of a two-species BJJ can be written as follows:
\begin{equation}
	\label{eq:quan1}
	\hat{H} = \hat{H}_1 + \hat{H}_2 + \hat{H}_{\rm 12},
\end{equation}
where
\begin{eqnarray}
	\label{eq:quan2}
	\hat{H}_1&=&\frac{u_1}{2}\left[(\hat{a}^{\dagger}_{L} \hat{a}_{L})^2 + (\hat{a}^{\dagger}_{R} \hat{a}_{R})^2 \right] -v_1\left(\hat{a}^{\dagger}_{L} \hat{a}_{R} + \hat{a}^{\dagger}_{R} \hat{a}_{L}\right), \nonumber \\
	\hat{H}_2&=&  \frac{u_2}{2}\left[(\hat{b}^{\dagger}_{L} \hat{b}_{L})^2 + (\hat{b}^{\dagger}_{R} \hat{b}_{R})^2\right]-v_2\left(\hat{b}^{\dagger}_{L} \hat{b}_{R} + \hat{b}^{\dagger}_{R} \hat{b}_{L}\right), \nonumber \\
	\hat{H}_{12} &=& u_{12} \left(\hat{a}^{\dagger}_{L} \hat{a}_{L} \hat{b}^{\dagger}_{L} \hat{b}_{L} + \hat{a}^{\dagger}_{R} \hat{a}_{R} \hat{b}^{\dagger}_{R} \hat{b}_{R}\right) \nonumber.
\end{eqnarray}
\begin{figure}[htbp!]
	\centering
	\includegraphics[width= 9.5 cm,height= 12 cm, keepaspectratio]{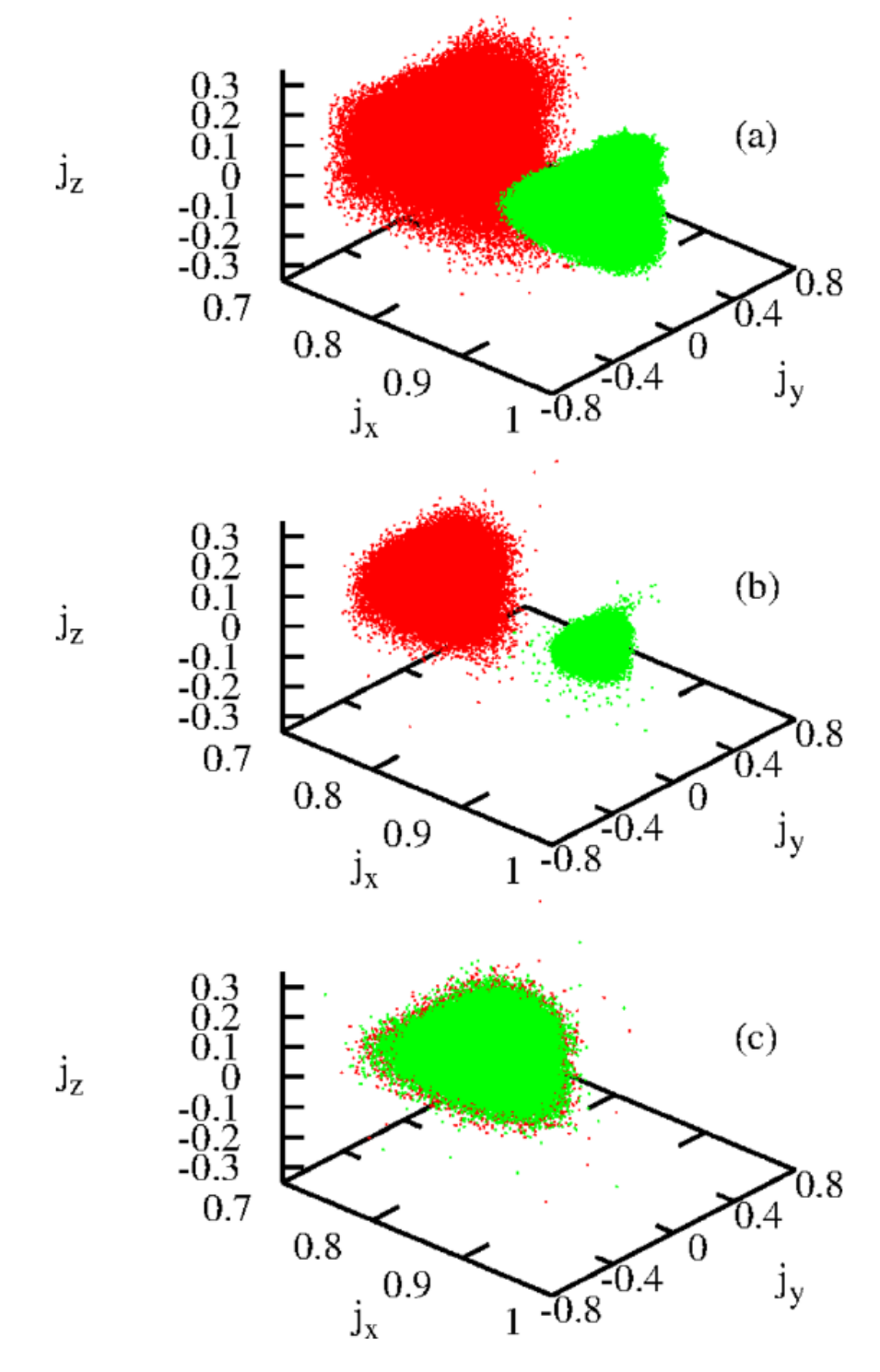}
	\caption{MS in many-body quantum system is distinguished by the region traversed by each subsystem during its evolution in the 3D space characterized by the average pseudoangular momentum $j_x = (2/N)\left\langle \hat{J}_x \right\rangle$, $j_y = (2/N)\left\langle \hat{J}_y \right\rangle$, and $j_z = (2/N)\left\langle \hat{J}_z \right\rangle$. The red and green colours correspond to species $a$ and $b$, respectively. Specifically, for the initial conditions $Z_1(0) = 0.2$, $Z_2(0) = 0.4$, and $N_1 = N_2 = 30$, we observe the following scenarios: (a) When $u_{12} = 0$ (desynchronization), (b) When $u_{12} = 0.008u$ (desynchronization), and (c) When $u_{12} = 0.5u$ (MS). In subplot (c), the two clouds representing the subsystems exhibit substantial overlap. This figure is taken from Qiu et al.~\citep{qiu14} with permission.} 
	\label{fig:quan}
\end{figure}
The Hamiltonian under consideration (Eq.~\ref{eq:quan1}) can be subjected to numerical diagonalization within the $N_d = (N_1+1)(N_2+1)$ dimensional space, defined by the tensor product of the many-body Fock basis $|N_{1,L}\rangle$ and $|N_{2,L}\rangle$ spanning the $a$ and $b$ Fock states, where $N_{1, L} = 0, 1, \cdots, N_1$ and $N_{2, L} = 0, 1, \cdots, N_2$. A general $N$-particle state can be expressed as 
\begin{equation}
	|\Psi\rangle = \sum_{N_{1, L} = 0}^{N_1} \sum_{N_{2, L} = 0}^{N_2} c_{N_{1, L}, N_{2, L}} | N_{1, L}, N_{2, L} \rangle
\end{equation}
The time-dependent Schr\"odinger equation determines the temporal evolution of any initial state. We can deduce the average particle numbers $\left\langle N_{1, \alpha}\right\rangle = \left\langle\Psi|\hat{a}^{\dagger}_{\alpha} \hat{a}_{\alpha}| \Psi \right\rangle$ and $\left\langle N_{2, \alpha}\right\rangle = \left\langle\Psi|\hat{b}^{\dagger}_{\alpha} \hat{b}_{\alpha}| \Psi \right\rangle$ (where $\alpha = L, R$) associated with the $L$ and $R$ modes upon computing the many-body state. The population imbalance for each species is quantified as $Z_{1(2)} = \left(N_{1(2), L} - N_{1(2), R}\right)/ N_{1(2)}$. To characterize the extent of condensation within each subsystem, $a$ and $b$, we employ the one-body density matrix $\rho$~\citep{julia10}. The traces of $\rho_1$ and $\rho_2$ are normalized to the number of atoms in each respective subsystem, denoted as $N_1$ and $N_2$.

We establish equivalent intraspecies interactions, specifically $u = u_1 = u_2$, with a value of $N u/v = 7.2$, and equivalent linear couplings, denoted as $v_1 = v_2 = v$. We utilize the Rabi time, $t_{Rabi} = \pi/v$, as our unit of time, and $\hbar/t_{Rabi}$ as our unit of energy. Our initial states consistently assume the form of coherent states for both species $a$ and $b$, where all atoms occupy the single-particle state $(1/\sqrt{2}) \left( \cos(\theta/2) \hat{a}^{\dagger}_{L} + \sin(\theta/2) \hat{a}^{\dagger}_{R}\right)$. The transition from desynchronization to MS state is graphically depicted in Fig.~\ref{fig:quan}. We plot the mean values of the pseudoangular momentum operators, $\hat{J}_x = (1/2) \left(\hat{a}^{\dagger}_{L} \hat{a}_{R} + \hat{a}_{L} \hat{a}^{\dagger}_{R}\right)$, $\hat{J}_y = (1/2) \left(\hat{a}^{\dagger}_{L} \hat{a}_{R} - \hat{a}_{L} \hat{a}^{\dagger}_{R}\right)$, and $\hat{J}_z = (1/2) \left(\hat{a}^{\dagger}_{L} \hat{a}_{L} - \hat{a}^{\dagger}_{R} \hat{a}_{R}\right)$, which can be readily constructed from the creation and annihilation operators of each species. As evidenced in the 3D representation, in the instances of desynchronized dynamics, i.e., when $u_{12} = 0$ (Fig.~\ref{fig:quan}a) and $u_{12} = 0.008u$ (Fig.~\ref{fig:quan}b), the domains explored by each subsystem in the $\left( \left\langle \hat{J}_x \right\rangle, \left\langle \hat{J}_y \right\rangle, \left\langle \hat{J}_z \right\rangle \right)$ space remain disjointed. In contrast, in the MS case (Fig.~\ref{fig:quan}c), both domains exhibit complete overlap. This characteristic can be interpreted as the many-body counterpart of the classical definition of MS, wherein the phase space domain covered by both subsystems coincides. To this end, a method for quantifying the transition from desynchronization to MS dynamics in classical systems involves examining the time-averaged bare energies of interacting subsystems. This energy-based index (Eq.~\ref{eq:bare_eng_bjj}) equally applies to studying the transition to MS in the quantum scenario.

MS has also been investigated in a pair of coupled Harper systems in classical and quantum contexts~\citep{sur20}. This study demonstrates the quantum counterpart of synchronization in a pair of coupled quantum-kicked Harper chains. Two spin chains are coupled through a time and site-dependent potential. The average interaction energy between the participating systems is used as an order parameter in classical and quantum contexts to establish a connection between the scenarios. In addition, this study also examines the entanglement between the chains and the difference between the average bare energies in the quantum context. Interestingly, all indicators suggest a connection between the MS transition in classical maps and a phase transition in quantum spin chains.

Zhang et al~\citep{zhang20} investigate the synchronization of two nonlinear mechanical modes of Bose-Einstein condensates in a closed quantum system, using classical and quantum measures. The aim is to reveal the macroscopic and microscopic properties of synchronized behaviours. The authors have used the Pearson correlation coefficient, orbital overlapping, and covering areas in the phase space based on mean-value dynamical equations to identify MS in the classical picture. On the other hand, the Husimi Q functions have been used to display the synchronized behaviours of quantum MS based on density overlapping and correlated probability dynamics in phase space. The quantum MS has been further investigated using two other quantum measures: the Mari measure and mutual information.

Another study~\citep{qiu22} explores MS in hybrid quantum-classical systems. The dynamics of the classical subsystem are modelled by Hamiltonian equations, while the Schr\"odinger equation governs the dynamics of the quantum subsystem. The authors have also explained the physical reason behind MS, which is the classical-like behavior of the quantum subsystem as the coupling strength increases. This effect leads to MS in quantum scenario mimicking MS in coupled classical Hamiltonian systems. Additionally, the frequency spectra of the MS states exhibit similar results to classical MS states with increasing coupling strength.

A similar kind of synchronization, termed hybrid synchronization (HS), has been studied considering the dynamics of coupled ultracold atomic clouds after changing the coupling strength between them~\citep{qiu15}. HS can coexist as compared to MS in the same model system. These two types of synchronizations differ in two ways. First, MS is a synchronization that occurs in the spatial coincidence sense and is characterized by a unique synchronization with an invariant measure in the phase space domain~\citep{hampton99}. Conversely, HS is a synchronization that occurs in the time coincidence sense and is a general term for synchronization that is characterized by strong time correlation in coupled quantum-classical dynamics. The time correlation in HS can be as strong as complete synchronization~\citep{pecora97} in coupled dissipative systems, while in terms of time correlation, MS is much weaker. Secondly, there are significant differences in setting the initial conditions and parameters for HS and MS. For HS, the quantum and classical subsystems must have initial conditions with exact correspondence. In contrast, the initial conditions cannot have exact correspondence and must be different in the case of MS.

\section{Conclusion and discussions}
\label{sec:con}
In conclusion, we have comprehended MS in interacting Hamiltonian systems in this paper. Each system shares a phase space domain with an identical invariant measure in the MS state. This kind of synchronization is observed when the Hamiltonian systems exhibit either quasiperiodic or chaotic dynamics. Although synchronization in dynamical systems has been a topic of intense research over the last few decades, discussions on MS needed to catch up. We have discussed MS incorporating both classical and quantum scenarios in this paper. Along with Hamiltonian systems with two interacting subsystems, studies on MS have also been extended to the many-body systems, i.e., Hamiltonian systems consisting of three or more subsystems. To this end, the notion of partial MS has been incorporated. Note that this kind of synchronization is relatively less robust than other types of synchronization observed in coupled dynamical systems. 

Besides, the anticipation of synchronization is an essential aspect of studying coupled dynamical systems~\citep{ghosh22, ghosh22_2}. It is well known that synchronization occurs when the coupling between the systems is strong enough to overcome the intrinsic differences in their dynamics. However, the anticipation of synchronization refers to predicting the conditions under which synchronization occurs before it happens. This can be accomplished through analytical, numerical, or experimental methods. Anticipation of synchronization is crucial for practical applications such as control and communication, where the ability to predict synchronization can aid in designing more effective control strategies or communication protocols. Moreover, the anticipation of synchronization can provide insights into the underlying mechanisms of collective behaviour in complex systems, leading to a better understanding of the dynamics of natural and engineered systems. MS can be anticipated using a parameter-aware reservoir computing technique based on machine learning~\citep{zhang22}.

Complex network~\citep{barabasi16} is another open direction for further research on MS. Various small-world networks represent different real-life systems. Furthermore, plenty of real-world complex systems are better represented by multi-layer networks and ignoring multiplexity with other layers may result in the wrong prediction of the outcome of coupled units in one layer. In multiplex, we have to deal with two types of coupling: inter-layer and intra-layer~\citep{bianconi18}. To date, one paper~\citep{tian23} focuses on this topic and investigates MS in a two-population network of coupled metronomes system. 

Finally, the investigation of indirectly coupled dynamical systems is of significant interest since it is a prevalent phenomenon in nature. Each dynamical system interacts with others through a common environment or medium such as air, water, and signalling chemicals. The emergence of collective behaviour in the natural world is closely related to indirect coupling. MS, as collective behaviour in Hamiltonian systems, can arise under various coupling schemes, including indirectly coupled schemes, as well as linearly and nonlinearly coupled schemes. An article~\citep{tian19} shows this is also the case for MS in coupled Hamiltonian systems. However, a lot needs to be uncovered in future to incorporate indirect coupling in the context of MS.

\section*{Acknowledgment}
The author thanks Dr. Tirth Shah and Dr. Saikat Sur for several fruitful discussions on measure synchronization. The author also thanks anonymous referees for their comments to improve the quality of the paper. This study is supported by the Czech Science Foundation, Project No. GA24-11113S and by the Czech Academy of Sciences, Praemium Academiae awarded to Dr. M. Palu\v{s}.

\bibliographystyle{plain}
\bibliography{template}

\end{document}